\documentclass[usenatbib]{mnras}

\usepackage{multirow}
\usepackage{rotating}
\usepackage{colortbl}
\usepackage{color}
\usepackage[fleqn]{amsmath}
\usepackage{amssymb}
\usepackage{amsfonts}
\usepackage{verbatim}
\usepackage{scalefnt}
\usepackage[percent]{overpic}
\usepackage[T1]{fontenc} 
\usepackage{aecompl} 
\usepackage{ulem}
\usepackage{cleveref}
\usepackage{xfrac}

\usepackage[dvipsnames]{xcolor}

\usepackage[percent]{overpic}

\def\msun{{\rm M}_{\sun}}
\def\lsim{\mathrel{\rlap{\lower 3pt \hbox{$\sim$}} \raise 2.0pt \hbox{$<$}}}
\def\gsim{\mathrel{\rlap{\lower 3pt \hbox{$\sim$}} \raise 2.0pt \hbox{$>$}}}

\newcommand{\comments}[1]{} 
\newcommand\T{\rule{0pt}{2.6ex}}       
\newcommand\B{\rule[-1.2ex]{0pt}{0pt}} 

\title[Generation of GWs and TDEs in clumpy galaxies]{Generation of gravitational waves and tidal disruptions in clumpy galaxies}

\author[B. Pestoni et al.]{Boris Pestoni,$^{1,2,3}$\thanks{E-mail: boris.pestoni@space.unibe.ch} Elisa Bortolas,$^{1}$ Pedro~R. Capelo$^{1}$ and Lucio Mayer$^{1,2}$\\
$^1$Center for Theoretical Astrophysics and Cosmology, Institute for Computational Science, University of Zurich,\\
Winterthurerstrasse 190, CH-8057 Z{\"u}rich, Switzerland\\
$^2$Physik-Institut, University of Zurich, Winterthurerstrasse 190, CH-8057 Z{\"u}rich, Switzerland\\
$^3$Physikalisches Institut, University of Bern, Sidlerstrasse 5, CH-3012 Bern, Switzerland}

\date{Accepted 2020 November 6. Received 2020 November 4; in original form 2020 July 24}

\pubyear{2020}

\begin{document}

\label{firstpage}

\pagerange{\pageref{firstpage}--\pageref{lastpage}}

\maketitle


\begin{abstract}
Obtaining a better understanding of intermediate-mass black holes (IMBHs) is crucial, as their properties could shed light on the origin and growth of their supermassive counterparts. Massive star-forming clumps, which are present in a large fraction of massive galaxies at $z \sim 1$--3, are among the venues wherein IMBHs could reside. We perform a series of Fokker--Planck simulations to explore the occurrence of tidal disruption (TD) and gravitational wave (GW) events about an IMBH in a massive star-forming clump, modelling the latter so that its mass ($10^8 \,\msun$) and effective radius ($100$~pc) are consistent with the properties of both observed and simulated clumps. We find that the TD and GW event rates are in the ranges of $10^{-6}$ to $10^{-5}$ and $10^{-8}$ to $10^{-7}$~yr$^{-1}$, respectively, depending on the assumptions for the initial inner density profile of the system ($\rho \propto r^{-2}$ or $\propto r^{-1}$) and the initial mass of the central IMBH ($10^5$ or $10^3\,\msun$). By integrating the GW event rate over $z = 1$--3, we expect that the Laser Interferometer Space Antenna will be able to detect $\sim$2 GW events per year coming from these massive clumps; the intrinsic rate of TD events from these systems amounts instead to a few $10^3$ per year, a fraction of which will be observable by e.g. the Square Kilometre Array and the Advanced Telescope for High Energy Astrophysics. In conclusion, our results support the idea that the forthcoming GW and electromagnetic facilities may have the unprecedented opportunity of unveiling the lurking population of IMBHs.
\end{abstract}



\section{Introduction}\label{sec:introduction}

Black holes (BHs) are typically categorized into two different species:  stellar-mass BHs \citep[SBHs; e.g.][]{Abbott_2016, Giesers_2017}, whose mass spans from  $\sim$10 to $\sim$100$\,\msun$, and supermassive BHs \citep[SMBHs; e.g.][]{Gillessen_2009, Event_Horizon_Telescope_Collaboration_2019}, with masses in the range of $10^6$--$10^{10}\,\msun$.

In the last few decades, observational astronomers have been pointing to an apparent dearth of BHs with masses between the heaviest SBHs ($\sim$100~$\msun$) and the lightest SMBHs ($\sim$10$^6\ \msun$) \citep[for a review, see][]{Mezcua_2017}: BHs inhabiting this loosely populated mass range have been named intermediate-mass BHs (IMBHs). While the lack of SBHs with masses above 100~$\msun$ can be justified within the current stellar evolutionary framework \citep{Mapelli2018},\footnote{We note, however, that primordial BHs with masses greater than 100~M$_{\sun}$ could in principle exist (but see e.g. \citealt{Carr_et_al_2020}, for recent observational constraints). Additionally, mergers of lower-mass BHs can in principle produce BHs with masses $>$100~M$_{\sun}$ \citep[][]{Abbott_2020}.} it is unclear whether a lower limit should exist in the  SMBH mass function, as the SMBH formation path is still largely debated \citep[e.g.][]{Volonteri2010,Valiante_et_al_2017}.

In fact, a number of compelling observations point to correlations between the SMBH mass and either the bulge \citep[e.g.][]{Kormendy2013} or the total \citep[e.g.][]{Reines_Volonteri_2015,Sahu_et_al_2019} stellar mass of the host; the extrapolation of such scaling relation to low-mass self-gravitating objects \citep[see also][]{Baldassare_et_al_2020}, such as dwarf galaxies and globular clusters, would imply IMBHs to be natural guests of these smaller astrophysical systems. Only recently, the first IMBH observations started to come forward \citep[e.g.][]{Gerssen_2002,Gebhardt_2005,Ulvestad_2007,Maccarone_2008,Noyola_2010,Lutzgendorf_kis_2012,Lutzgendorf_ki_2013,Lanzoni_2013,Haggard_2013,Reines_et_al_2013,Casares_2014,Baldassare2015,Kiziltan_2017,Perera_2017,Lin2018,Mezcua_et_al_2018} and many of them remain debated. It is unclear whether the paucity of IMBH detections has to be interpreted as an intrinsic lack of these objects, or rather observational biases render their detection inherently more challenging \citep[e.g.][]{Mezcua_2017}.

The idea that IMBHs could inhabit the centre of globular clusters dates back to the late 1960s  \citep[][]{Wheeler_1968,Wyller_1970}; from that moment on, a number of theoretical studies focused on  the formation path of IMBHs within their hosts. For instance, \citet{Miller_2002} proposed the formation of an IMBH starting from a massive SBH that increases its mass thanks to the coalescence with smaller SBHs. Alternatively, a series of collisions among massive stars,  enhanced by the segregation of these bodies at the centre of their hosts system and the resulting increase in the central density, could lead to the growth of a supermassive object in a runaway fashion \citep[e.g.][]{Portegies-Zwart2002, Portegies-Zwart2004}, possibly aided by the presence of gas \citep{Reinoso2020}. Additionally, \citet{Giersz_2015} suggested that an IMBH could result from dynamical interactions of hard binaries containing an SBH with other bodies in the globular cluster. A significant number of the proposed formation scenarios rely on dynamical processes that bring to a series of collisions within the host system. For this, very dense stellar  environments subject to mass segregation, core collapse, and binary interactions seem to be ideal environments for the formation of IMBHs \citep[e.g.][]{Gurkan_fr_2004,Gultekin_2004,Gurkan_2006,Freitag_2006}.

In the near future, thanks to new observatories such as the Laser Interferometer Space Antenna \citep[LISA;][]{LISA_2017,Barack_et_al_2019}, TianQin \citep{Luo2016}, the Square Kilometre Array \citep[SKA;][]{Dewdney_et_al_2009}, the extended ROentgen Survey with an Imaging Telescope Array \citep[eROSITA;][]{eROSITA_2012}, and the Advanced Telescope for High Energy Astrophysics \citep[Athena;][]{Barcons2012}, we will have the capability to observe  IMBHs via new strategies, detecting the signal that these objects may generate as a consequence of the interaction with their surroundings up to the relatively high redshift Universe. Such events could manifest themselves in the form of both gravitational wave (GW) signals -- in particular from extreme and intermediate mass-ratio inspirals \citep[EMRIs and IMRIs; e.g.][]{Amaro-Seoane_2007} -- and tidal disruption events \citep[TDEs; e.g.][]{Kashiyama_2016}.\footnote{Note that TDEs are also expected to produce a GW background, that is in principle also accessible to forthcoming GW observatories up to $z\approx3$ \citep[][]{Toscani_et_al_2020}.} The search for this kind of events about IMBHs is crucial, considering that these objects could constitute the high-redshift seeds for the SMBHs we observe at the present day; thus, the aforementioned detections could provide new crucial insights on the cosmic origin and growth of SMBHs \citep[][]{Volonteri2010}.

Recent observations have shown that a large fraction \citep[$\gtrsim$50 per cent;][]{Shibuya_et_al_2016} of massive galaxies in the redshift range $z \sim 1$--3 present massive star-forming clumps, with typical masses of $\sim$10$^7$--$10^8$~M$_{\sun}$ and, in some cases, even $\sim$10$^9$~M$_{\sun}$ \citep[e.g.][]{Elmegreen_2009}. These clumps are likely to constitute a privileged setting for the formation of IMBHs and, consequently, for the occurrence of IMRIs and TDEs that could be potentially detected by the forthcoming facilities. The existence of such numerous massive clumps in many galaxies at the peak of star formation and galaxy merger rates is corroborated by both spatially resolved observations \citep[e.g.][]{Huertas-Company2020} and numerical simulations \citep[e.g.][]{Tamburello_et_al_2015}.

For this reason, in this paper we adopt a numerical approach to estimate the number of absorption events\footnote{In this work, we define absorption an event in which a stellar-mass object (star or compact object) is `absorbed' by the central IMBH. We then distinguish between TDEs and GW events, which can be subsequently divided into EMRIs, IMRIs, or plunges, depending on the mass ratio and on the number of orbits of the stellar-mass compact object around the IMBH.} occurring in a typical massive stellar clump\footnote{An analysis concerning the possibility of generating absorption events in globular clusters, common also at $z=0$, has been carried out by \citet{Fragione_Ginsburg_et_al_2018,Fragione_Leigh_et_al_2018}.}, whose properties are extrapolated from the isolated simulation of a star-forming galaxy by \citet{Tamburello_et_al_2015}, in the assumption that an IMBH is found within the host system.

The paper is organized as follows: in Section~\ref{sec:methods}, we describe the numerous aspects of our model and the numerical set-up; in Section~\ref{sec:results}, we present our results; in Section~\ref{sec:discussion}, we discuss our findings; and we conclude in Section~\ref{sec:conclusion}.


\section{Methods}\label{sec:methods}

In order to estimate the rates of absorption events occurring in stellar clumps, we first obtain the characteristic parameters of a typical clump from an isolated simulation of a star-forming galaxy (Section~\ref{sec:initial_cond}). We then use such quantities to build analytical models with different density profiles (Section~\ref{sec:profile_models}), which are used as the initial conditions adopted in the Fokker--Planck simulations (see Section~\ref{sec:fokker_planck_simul}) performed to keep track of the event rates.


\subsection{Initial conditions}\label{sec:initial_cond}

In order to build a realistic model for the star-forming clumps in which IMBHs may reside, we rely on the numerical work of \citet{Tamburello_et_al_2015}, where the properties of massive star-forming clumps in isolated galaxies were studied in detail. They performed a large suite of galaxy simulations, to assess the dependence of gas fragmentation properties in a galaxy on sub-grid physics, galaxy mass, structural parameters, and resolution, using the smoothed particle hydrodynamic code {\fontfamily{qcr}\selectfont gasoline2} \citep[][]{Wadsley_et_al_2017}, an updated version of {\fontfamily{qcr}\selectfont gasoline} \citep{Stadel_2001,Wadsley_2004}. In this paper, we specifically analyse the clumps forming in their run~27 \citep[see tables~1 and 2 of][]{Tamburello_et_al_2015}, in which they simulated a dark matter halo with an embedded stellar and gaseous disc, with a virial and a baryonic mass of $2.5 \times 10^{12}$ and $7.8 \times 10^{10}$~M$_{\sun}$, respectively, a gas fraction of 50 per cent, and a virial concentration of 6. This choice of parameters was particularly conducive to the formation of long-lasting massive clumps. Star formation and stellar (blast-wave) feedback were modelled following the prescriptions of \citet{Stinson_et_al_2006}, with a star formation efficiency, threshold temperature, and threshold density of 0.01, $3 \times 10^4$~K, and 10~$m_{\rm H}$~g~cm$^{-3}$, respectively, and an energy injection of $4 \times 10^{50}$~erg per supernova explosion. The initial numbers of particles were $1.2 \times 10^6$, $10^5$, and $10^5$, for dark matter, gas, and stars, respectively, whereas the softening, identical for all particles, was set equal to 100~pc.

Using {\fontfamily{qcr}\selectfont skid} \citep[][]{Stadel_2001}, a group finder for $N$-body simulations, we extracted the mass and half-mass radius of the largest clumps formed in the simulation, finding a typical stellar clump mass

\begin{equation}
\label{eq:typical_total_mass}
M_{\rm clus} \approx 10^8\msun{},
\end{equation}

\noindent compatible with what was observationally found by \citet{Elmegreen_2009} in their Hubble Ultra-Deep Field analysis of clump-cluster and chain galaxies, and a characteristic half-mass radius

\begin{equation}
\label{eq:typical_r_hm}
r_{\rm hm} \approx 100\ \mathrm{pc}.
\end{equation}

We caution that the mass and length resolution in the simulation by \citet{Tamburello_et_al_2015} are too coarse to perform a self-consistent modelling of the clumps starting from the density profile obtained from the simulation; for this reason, we create analytical models using equations~(\ref{eq:typical_total_mass}) and (\ref{eq:typical_r_hm}) as constraints, as described in Section~\ref{sec:profile_models}.


\subsubsection{Composition and structure of the cluster}\label{sec:profile_models}

\begin{table*}
\centering
\begin{tabular}{cccccc}
$m_{\rm in}$ (i) & Species (ii) & Number frac. (iii) & Object mass (M$_{\sun}$)  (iv) & Object radius (R$_{\sun}$) (v) & Num. of objects (vi) \T \B \\
\hline
$0.01 \leq m_\star/\mathrm{M}_{\sun} < 1$ & MS & $0.982$ & 0.4, \textbf{0.6}, 0.8 & 0.4, \textbf{0.6}, 0.8 & $1.575 \times 10^8$  \T \B \\
$1 \leq m_\star/\mathrm{M}_{\sun} < 8$ & WD & $1.44\times10^{-2}$ & 0.7, \textbf{1.0}, 1.3 & 0.012, \textbf{0.008}, 0.004 & $2.305 \times 10^6$ \T \B \\
$8 \leq m_\star/\mathrm{M}_{\sun} < 30$ & NS & $3.36\times10^{-3}$ & 1.6, \textbf{2.3}, 3.0 & --- & $5.392 \times 10^5$ \T \B \\
$30 \leq m_\star/\mathrm{M}_{\sun} < 120$ & SBH & $6.13\times10^{-4}$ & 10, \textbf{20}, 30 & --- & $9.834 \times 10^4$ \T \B \\
\hline
\end{tabular}
\caption{Parameters associated with the four stellar-remnant families -- main-sequence stars (MSs), white dwarfs (WDs), neutron stars (NSs), and SBHs -- included in our runs: (i) initial mass range, (ii) name of the stellar-remnant species, (iii) final number fraction associated with each species (assuming a 1-Gyr-old stellar population), (iv) mass of each single object in the family (we explored three values; the default is in bold), (v) respective stellar radius (for MSs and WDs), and (vi) number of objects. The total number of objects is ${1.6} \times 10^8$ for all cases (meaning that the total stellar mass can vary by up to $\sim$30 per cent when we change the object mass; see Section~\ref{sec:other_changes}).}
\label{tab:evolved_kroupa_mf_to_nf}
\end{table*}

For simplicity, we assume that the entire gaseous mass residing in the clusters described in \citet{Tamburello_et_al_2015}, which is of the order of the stellar mass, collapses in stars and/or is evaporated from the clump by means of stellar/BH feedback.\footnote{Depending on the mechanisms at play, the final stellar mass can vary by a factor of at most 2. Therefore, for the computations of the default case, we will assume that $M_{\rm clus} = 10^8\msun{}$. We also assume $r_{\rm hm} = 100\ \mathrm{pc}$ for all our scenarios.}

Such assumption allows us to model the cluster as a dissipationless system composed only by stellar objects. Specifically, we assume the cluster to initially follow a spherical density profile belonging to the  \citet{Dehnen_1993} family \citep[see also][]{Tremaine_1994}:

\begin{equation}
\label{eq:dehnen_model}
\rho_{\rm D}(r)=\frac{(3-\gamma)M_{\rm clus}}{4\pi}\frac{a}{r^\gamma(r+a)^{4-\gamma}},
\end{equation}

\noindent where $a$ is the scale radius of the system, $M_{\rm clus}$ its total mass, and $\gamma\in[0;3)$ its logarithmic inner density slope. The half-mass radius of such systems is related to the scale radius via the relation

\begin{equation}
\label{eq:dehnen_a}
r_{\rm hm,D}=\frac{a}{2^{1/(3-\gamma)}-1}.
\end{equation}

In this work, we only model two cases: a \citet{Jaffe_1983} profile ($\gamma=2$, default case) and a \citet{Hernquist_1990} profile ($\gamma=1$). The choice of the default configuration ($\gamma = 2$) is motivated by the fact that violent relaxation and dissipational collapse could produce a system with a density slope $\gamma \approx 2$, as hinted by past studies \citep[see e.g.][]{Hjorth1991, Treu2002}; we additionally modelled the cluster with a \citeauthor{Hernquist_1990} density profile, to assess the dependence of our results on the choice of the initial profile (see Section~\ref{sec:change_in_the_density_profile}).

We assume that the clump is composed of one central IMBH (see below) and four stellar families, each  following the same density profile at the beginning of the integration: MSs, WDs, NSs, and SBHs. The Fokker--Planck treatment, and in particular the {\fontfamily{qcr}\selectfont phaseflow} code described below (see Section~\ref{sec:fokker_planck_simul}) can only handle a finite number of components, each having a fixed mass. Hence, we could not adopt a smooth and non-discretized mass function, but assigned instead a typical mass to each of the four components, whose values are listed in Table~\ref{tab:evolved_kroupa_mf_to_nf}. Such values are derived assuming that stars were born with a \citet{Kroupa_2001} initial mass function and the stellar population is 1~Gyr old (i.e. the cluster is assumed to have a turnoff mass of 2.512~M$_{\sun}$), which is the typical time-scale within which massive clumps are expected to dissolve in the tidal field of their host galaxy \citep[][]{Tamburello_et_al_2015}.

In particular, we adopt the initial-to-final mass relation proposed by \citet{Merritt_2013}. The mass of each object in a given species is shown in Table~\ref{tab:evolved_kroupa_mf_to_nf}: for each component, we selected a default mass value (in bold in the table) and performed additional runs varying it (as described in Section~\ref{sec:other_changes}). Combining the object masses with equation~\eqref{eq:typical_total_mass}, we obtain the numbers of objects in the default configuration, as shown in Table~\ref{tab:evolved_kroupa_mf_to_nf}.

We assume that an IMBH of mass $m_\bullet$ exists at the centre of the massive clump. There are multiple ways to achieve this in Nature. In particular, the IMBH could be either formed \textit{in situ} via runaway collapse of relatively massive stars, or be captured by the massive stellar clump. We speculate on the possible channels of \textit{in situ} formation or capture in Section~\ref{sec:discussion}. Here we assume that an IMBH with an initial mass $m_{\bullet0} = 10^5$~M$_{\sun}$ lies at equilibrium at the centre of the system; this mass choice implies that the cluster and its IMBH sit on the extrapolated bulge mass--BH mass relation observed in more massive systems in the local Universe \citep{Kormendy2013}. We additionally performed a test with $m_{\bullet0} = 10^3$~M$_{\sun}$, to assess how our results change when we vary the initial IMBH mass (see Section~\ref{sec:imbh_mass_change_variat}).

\subsubsection{Absorption events}\label{sec:TDE_and_IMRI}

Our main aim is to obtain the rates of TDEs and GW events occurring in the modelled stellar clump hosting an IMBH.

A TDE occurs when a MS or a WD (when we assume that WDs can be tidally disrupted; see Section~\ref{sec:other_changes}) approaches the close vicinity of the IMBH and is tidally disrupted as the IMBH gravitational attraction overcomes the self gravity of the stellar object. The threshold separation for a TDE to occur is named the tidal radius, and depends on the stellar mass, $m_\star$, radius, $r_\star$, and internal structure, scaling as \citep[][]{Hills_1975, Rees_1988}

\begin{equation}
\label{eq:tidal_radius}
\begin{split}
r_{\rm tid}& = \eta^{2/3} \left(\frac{m_\bullet} {m_\star}\right)^{1/3}r_\star
 \approx 7.4\times 10^{-7}\left(\frac{m_{\bullet}}{10^5~\msun{}}\right)^{1/3} \times \\ & \times \left(\frac{m_\star}{0.6~\msun{}}\right)^{-1/3}\left(\frac{r_\star}{0.6~\mathrm{R}_{\sun}}\right) \mathrm{pc},
\end{split}
\end{equation}

\noindent where $\eta$ is a parameter of order unity that depends on the mass and equation of state of the stellar object \citep[][]{Chandrasekhar_1939,Shapiro_Teukolsky_1983}, and which we set equal to 1. The radii for MSs and WDs used in our runs are listed in Table~\ref{tab:evolved_kroupa_mf_to_nf}: we extrapolated the MS radii from the empirical mass--radius relation in \citet{Chen_2014}, whereas we used the relationship in \citet{Magano_2017} for the WD radii.

If the stellar object is too compact, it is expected to enter the IMBH horizon prior to undergoing a TDE. In fact, a gravitational capture, and the possibly associated EMRI or IMRI (depending on the mass ratio),\footnote{In this work, we assume the inspiral's mass ratio to be extreme (EMRI), intermediate (IMRI), or moderate, when the mass ratio of the two objects is $q < 10^{-4}$, $10^{-2} > q \ge 10^{-4}$, and $q \le 10^{-2}$, respectively. The choice of these mass-ratio thresholds does not affect the results.} occurs when a stellar-mass compact object (such as an SBH, an NS, or a WD -- if the latter is not tidally disrupted) finds itself on an orbit that brings it inside the IMBH event horizon. The capture radius can be approximated as

\begin{equation}
\label{eq:capt_radius_compact_3}
r_{\rm cap}= \alpha\frac{Gm_{\bullet}}{c^2} \approx4.8\times 10^{-9} {\alpha}  \left(\frac{m_{\bullet}}{10^5\;\msun{}}\right)\mathrm{pc},
\end{equation}

\noindent where $G$ is the gravitational constant, $c$ is the speed of light in vacuum, and $\alpha$ depends on the orbital properties of the inspiralling object and on the IMBH spin.\footnote{We do not consider the effect of the inspiralling compact object's spin, as it is assumed to be negligible (e.g. \citealt{Barack_et_al_2004}; however, see \citealt{Han_et_al_2010}).} If we consider a compact object inspiralling on a nearly circular orbit about a non-spinning IMBH, then $\alpha = 6$, so that $r_{\rm cap}$ equals the radius of the IMBH innermost stable circular orbit (ISCO). However, the majority of compact objects are expected to reach the IMBH vicinity via two-body scatterings with other stars, that settle them on nearly radial orbits \citep[e.g.][]{Hils1995, Bortolas2019}; if such orbits do not circularise due to dynamical friction or GW emission, the critical $\alpha$ for a capture to occur is $\alpha = 8$, in the assumption of a non-spinning central IMBH \citep[][]{Will_2012}. The aforementioned parameter takes different values if the central IMBH is spinning: in particular, $\alpha =2 $ $(11.6)$ for a compact object inspiralling on a non-circular, prograde (retrograde) orbit on to a maximally spinning IMBH \citep[][]{Will_2012}.

While all compact objects crossing $r_{\rm cap}$ are expected to be accreted on to the central BH, not all of them can produce a {\it detectable} GW event. In fact, it is common use to distinguish between direct plunges and inspirals. In the former case, the compact object is directly scattered on to $r_{\rm cap}$; in the latter case, the compact object gradually inspirals on to the IMBH via emission of GWs, possibly producing a long-lived signal that can be detected by forthcoming observatories such as LISA. The Fokker--Planck approach does not allow us to distinguish between EMRIs/IMRIs and direct plunges; for simplicity, we will refer to both   as GW events, and we will comment upon the rates of both events in Section~\ref{sec:discussion}.

In what follows, we  refer to both the tidal or capture radius (depending on the species) as the absorption radius, $r_{\rm abs}$.


\subsection{Fokker--Planck simulation}\label{sec:fokker_planck_simul}

\begin{figure}
\centering
\includegraphics[width=0.4\textwidth]{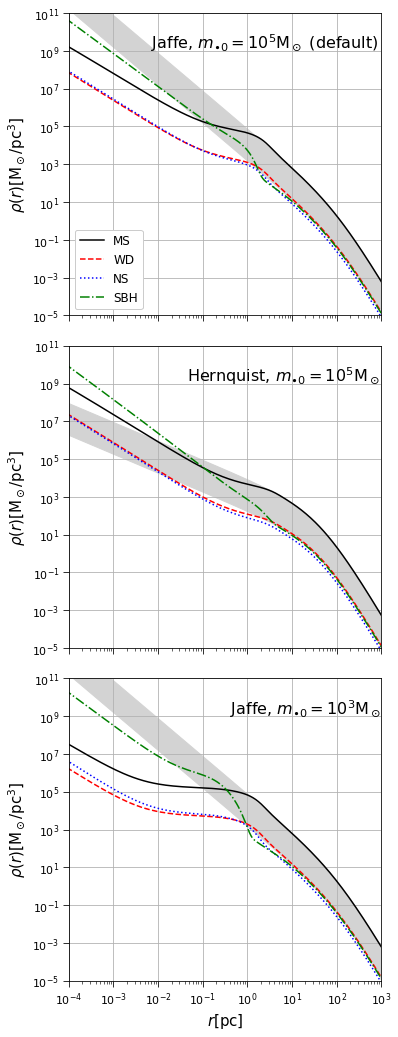}
\caption{Density profiles of MSs (black solid line), WDs (red dashed line), NSs (blue dotted line), and SBHs (green dash-dotted line), at 1~Gyr for the default configuration (top panel; \citeauthor{Jaffe_1983} model with $m_{\bullet 0} = 10^5$~M$_{\sun}$), for the case in which we adopted a \citeauthor{Hernquist_1990} density profile (central panel), and for the case in which the central IMBH has an initial mass of $10^3 \msun{}$ (bottom panel). The grey-shaded regions show the initial density profile of all species (with MSs and SBHs having the highest and lowest density, respectively).} 
\label{fig:density_1gyr_mf_to_nf}
\end{figure}

The number of GW events and TDEs occurring in the stellar clump is obtained via the {\fontfamily{qcr}\selectfont phaseflow} code \citep{Vasiliev_2017}, which is part of the publicly available {\fontfamily{qcr}\selectfont agama} library \citep{Vasiliev_2019}. {\fontfamily{qcr}\selectfont phaseflow} is a Fokker--Planck code that computes the dynamical evolution of a spherical isotropic system by solving the coupled system of Poisson and orbit-averaged Fokker--Planck 1D equations for the gravitational potential, the density, and the distribution function; note that {\fontfamily{qcr}\selectfont phaseflow} is the first code adopting the phase volume instead of the energy as the argument for the distribution function, thus rendering the code extremely efficient. {\fontfamily{qcr}\selectfont phaseflow} can simultaneously handle several species of stars (components), each with its own distribution function. In addition, it  can account for the presence of a central BH, and it incorporates a sink term that can mimic the loss of stars on to the central BH, thus delivering the rates of TDEs and GW events.\footnote{Note that the description of such events in terms of orbit-averaged Fokker--Planck equation is intrinsically approximated.} {\fontfamily{qcr}\selectfont phaseflow} has been successfully tested in the context of e.g. nuclear star clusters \citep[][]{Generozov_et_al_2018,Emami_Loeb_2020}, \citet{BahcallWolf1976} cusps \citep[][]{Vasiliev_2017}, primordial BH mass functions \citep[][]{Zhu_et_al_2018,Stegmann_et_al_2020}, and TDE rates in galaxy merger remnants and nuclear star clusters \citep[][]{Pfister_et_al_2019,Pfister_et_al_2020}.

In this study, the IMBH is incorporated in the simulation assuming it is at equilibrium with the surrounding system since the beginning of the integration, and the initial distribution function is constructed in the combined potential of the IMBH and the stellar-mass objects. While this is of course an approximation, violent relaxation that could have acted in the system at the stages of violent \textit{in situ} star formation and IMBH formation may have brought the system to dynamical relaxation even if its lifespan is shorter than its relaxation time.


\subsubsection{Simulation parameters}\label{sec:simul_params}

\begin{figure*}
\centering
\includegraphics[width=1.0\textwidth]{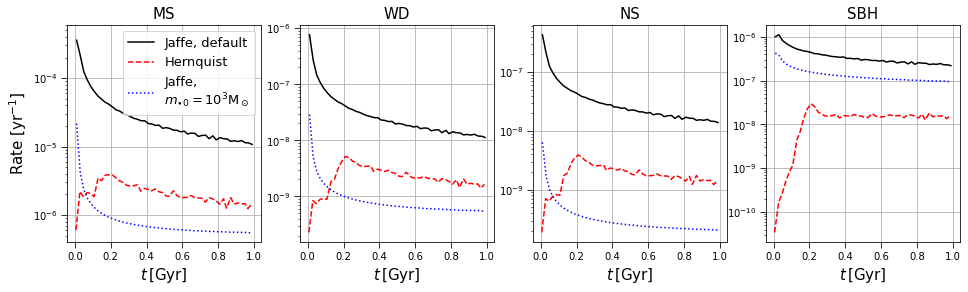}
\caption{Rate of absorption events (TDEs for MSs and WDs; GW events for NSs and SBHs) as a function of time occurring in the default configuration (black solid line; see Section~\ref{sec:default_scenario}), in the case adopting a \citeauthor{Hernquist_1990} instead of a \citeauthor{Jaffe_1983} initial density profile (red dashed line; see Section~\ref{sec:change_in_the_density_profile}), and in the case assuming an IMBH with initial mass of $10^3 \msun{}$ instead of $10^5 \msun{}$ (blue dotted line; see Section~\ref{sec:imbh_mass_change_variat}). The event rates associated with the default case are typically larger than those obtained in the other two scenarios by at least one order of magnitude, except for the case of SBHs.} 
\label{fig:event_rate_init_conf_mf_to_nf}
\end{figure*}

In this section, we describe the parameters of our simulations, focusing first on the default configuration.

We account for the presence of four species of stellar-mass objects -- MSs, WDs, NSs, and SBHs -- each distributed according to the same density profile: a \citeauthor{Jaffe_1983} profile (see the grey-shaded region in the top panel of Fig.~\ref{fig:density_1gyr_mf_to_nf}) with total mass of $10^8$~M$_{\sun}$ and half-mass radius $r_{\rm hm,D} = 100$~pc. As mentioned in Section~\ref{sec:profile_models}, the number fraction assigned to each stellar species and the mass of each single object are displayed in Table~\ref{tab:evolved_kroupa_mf_to_nf}.

The central IMBH is given an initial mass $m_{\bullet0} = 10^5$~M$_{\sun}$, and its mass increases via TDEs and GW events as the system evolves, the `absorption' boundary for such events being set according to equations~\eqref{eq:tidal_radius} and \eqref{eq:capt_radius_compact_3} (with $\alpha=8$), respectively; in both cases, this radius increases during the simulation as the IMBH mass grows. We also assume that WDs get tidally disrupted rather than undergoing an IMRI event, in agreement with, e.g. \citet{Haas_2012} and \citet{Anninos_2018}. 

When a stellar object (MS or WD) gets tidally disrupted, 30 per cent\footnote{We varied this fraction between 10 and 50 per cent and found that the results are not affected; note that, when a TDE occurs, the mass that is not added to the central IMBH is not included in the simulation from that moment on, thus the total mass is not conserved throughout the integration. } of its mass is accreted on to the IMBH, whereas the whole compact object's mass is added to the IMBH when a GW event occurs.

We set the Coulomb logarithm, $\ln\Lambda$, which tunes the significance of two-body relaxation in the Fokker--Planck code, equal to 16, since $\ln\Lambda \approx \ln \left({r_{\rm hm}}/{b_{90}}\right)$, where $b_{90}\sim 10^{-5}$ pc is the $90^{\circ}$ deflection radius for stellar encounters.

Finally, we initialized the Fokker--Planck solver so that the distribution function is computed on a grid in phase volume with 400 logarithmically spaced points, and we chose the quadratic discretization method \citep[method 2 in {\fontfamily{qcr}\selectfont phaseflow};][]{Vasiliev_2017} and an accuracy parameter for the time integration of $10^{-2}$. We checked that our results do not depend on the choice of such parameters. We evolved each realization for 1~Gyr, the typical massive-clump lifetime \citep[][]{Tamburello_et_al_2015}.


\section{Results}\label{sec:results}

In this section, we first present the rate of absorption events when considering the default scenario (Section~\ref{sec:default_scenario}) and then show how the results change when we vary the physical parameters of our simulation (Section~\ref{sec:dependence_on_parameters}).


\subsection{Default scenario}\label{sec:default_scenario}

Fig.~\ref{fig:event_rate_init_conf_mf_to_nf} shows the rate of absorption events (TDEs for MSs and WDs; GW events for NSs and SBHs) as a function of time for the default case (together with other scenarios discussed below), whereas Table~\ref{tab:event_rates_init_conf_mf_to_nf} lists the average rates of events per year and the relative number fraction of objects belonging to each species captured or disrupted by the IMBH. Overall, the highest event rates in the default scenario are associated with MSs ($3.7\times 10^{4}$ TDEs per Gyr), followed by SBHs ($3.7\times 10^{2}$ IMRIs per Gyr), WDs (49 TDEs per Gyr), and NSs ($42$ EMRIs per Gyr). This reflects the fact that MSs are the most abundant species in the system (Table~\ref{tab:evolved_kroupa_mf_to_nf}). The absorption event rate of SBHs is larger than that of WDs and NSs, despite the former species having a lower number fraction than the latter two. This can be understood in terms of mass segregation: SBHs have the largest mass among the four components by a large margin, hence they sink towards the centre of the system while pushing the lighter objects at larger distances from the centre. This fact can be appreciated by looking at the inner region in the top panel of Fig.~\ref{fig:density_1gyr_mf_to_nf}, where the SBH profile has initially the lowest normalization and, at the end of the integration, ends up with the largest.

\begin{table*}
\centering
\begin{tabular}{ccccc}
\multicolumn{2}{c}{} & \citeauthor{Jaffe_1983}, $m_{\bullet0} = 10^5$~M$_{\sun}$ & \citeauthor{Hernquist_1990}, $m_{\bullet0} = 10^5$~M$_{\sun}$ & \citeauthor{Jaffe_1983}, $m_{\bullet0} = 10^3$~M$_{\sun}$ \T \B \\
\hline
\multirow{2}{*}{MS} & Rate ($\mathrm{yr}^{-1}$) & $3.7\times 10^{-5}$ & $2.2\times 10^{-6}$ & $1.2\times 10^{-6}$ \T \B \\
& Accreted fraction & $2.4 \times 10^{-4}$ & $1.4 \times 10^{-5}$ & $7.9\times 10^{-6}$ \T \B \\
\hline
\multirow{2}{*}{WD} & Rate ($\mathrm{yr}^{-1}$) & $4.9\times 10^{-8}$ & $2.3\times 10^{-9}$ & $1.5\times 10^{-9}$ \T \B \\
& Accreted fraction & $2.1 \times 10^{-5}$ & $1.0 \times 10^{-6}$ & $6.4\times 10^{-7}$ \T \B \\
\hline
\multirow{2}{*}{NS} & Rate ($\mathrm{yr}^{-1}$) & $4.2\times10^{-8}$ & $1.9\times10^{-9}$ & $4.6\times 10^{-10}$ \T \B \\
& Accreted fraction & $7.8 \times 10^{-5}$ & $3.5 \times 10^{-6}$ & $8.5\times 10^{-7}$ \T \B \\
\hline
\multirow{2}{*}{SBH} & Rate ($\mathrm{yr}^{-1}$) & $3.7\times10^{-7}$ & $1.4\times10^{-8}$ & $1.4\times 10^{-7}$ \T \B \\
& Accreted fraction & $3.8 \times 10^{-3}$ & $1.4 \times 10^{-4}$ & $1.4\times 10^{-3}$ \T \B \\
\hline
\end{tabular}
\caption{For each of the four stellar-mass species, we list the TDE (for MSs and WDs) or GW event (for NSs and SBHs) rates and the number fraction of objects in each family that are accreted on to the IMBH by the end of the integration, with respect to the initial number of bodies of a given species inside the cluster. We list the numbers for the default case (\citeauthor{Jaffe_1983}, $m_{\bullet0} = 10^5$~M$_{\sun}$) and the comparison cases in which we use either a \citeauthor{Hernquist_1990} profile or a central IMBH with $m_{\bullet0} = 10^3$~M$_{\sun}$.
}
\label{tab:event_rates_init_conf_mf_to_nf}
\end{table*}

The SBH mass segregation can occur over $\sim$0.2 relaxation times either in the strong or in the weak (standard) regime \citep{Preto2010,Merritt_2013}. In the weak (standard) regime, SBHs are numerous enough at all radii within the IMBH influence sphere (defined as the region enclosing twice the IMBH mass) to dominate the relaxation rate, and they develop a so-called \citeauthor{BahcallWolf1976} cusp, with a logarithmic density slope $\rho \propto r^{-7/4}$. In the strong regime, a range of radii within the IMBH influence sphere exists where SBHs do not dominate the density profile, even after segregation occurred; in that outer region, they settle in a steeper cusp, $\rho\propto r^{-11/4}$ \citep{Alexander_2009, Amaro-Seoane2011}, while retaining the standard \citeauthor{BahcallWolf1976} cusp at the small scales where they keep dominating the relaxation rate. The other, lower-mass components typically settle into a shallower cusp (with roughly $\rho\propto r^{-\alpha}$, $\alpha=1.4$--1.5) \citep[e.g.][]{Merritt_2013}. The two segregation scenarios can be discerned via the parameter

\begin{equation}
\label{eq:delta_def}
\Delta = \frac{4\rho_{\rm H} m_{\star{\rm H}}}{(3+m_{\star{\rm H}}/m_{\star{\rm L}}) \rho_{\rm L}m_{\star{\rm L}}},
\end{equation}

\noindent where $\rho_{\rm H}$ is the mass density of heavy bodies (i.e. SBHs), $\rho_{\rm L}$ that of light bodies (i.e. MSs; we neglect here WDs and NSs, as they are relatively rare), and, in this particular case, $m_{\star{\rm H}} = 20$~M$_{\sun}$ and $m_{\star{\rm L}} = 0.6$~M$_{\sun}$ are the masses of a single body of the two populations; if $\Delta \ll 1$ ($\gg$1), the system is in the strong (weak) segregation regime. In our system, $\Delta\gtrsim1$ within the innermost parsec, implying that the strong regime is never in place. The same  applies to the other systems described in the next sections. 

The rate of TDEs and GW events is not constant along the simulation, as displayed in Fig.~\ref{fig:event_rate_init_conf_mf_to_nf}. In the first $\lesssim$0.1~Myr, the rate of events can be orders of magnitude larger than that at later times. In fact, within the IMBH influence sphere, the system  almost immediately develops a \citeauthor{BahcallWolf1976} cusp in the SBHs and a profile that scales as $\rho\propto r^{-1.45}$ in the other components, resulting in a drop of the density profile of all species. After the cusp has developed,  the system further slowly evolves, expanding in a self-similar fashion and producing a slow drop of the inner density normalization. This expansion is powered by the energy source of the IMBH, which absorbs stars with large negative energy, thus acting as a heat source, as discussed in detail in \citet{Vasiliev_2017}. The system expansion and associated drop in the central density produce the slow decrease in all the event rates at late times.
If one compares the late event rate for any species (from Fig.~\ref{fig:event_rate_init_conf_mf_to_nf}) with the average rate shown in Table~\ref{tab:event_rates_init_conf_mf_to_nf}, the values differ at most by factors of a few. This means that, even neglecting the initial burst of events, our results remain valid at least at the order-of-magnitude level. 

The IMBH influence radius is $\approx 0.2$ pc at the beginning of the simulation and reaches $\approx 0.9$ pc by the end of the integration; the increase of the influence radius is mirrored by the decrease in the central density following the expansion, as can be inferred from Fig.~\ref{fig:density_1gyr_mf_to_nf}.

Finally, it is worth mentioning that the fractional mass accreted by the central IMBH via TDEs and GW events by the end of the integration is $\sim$0.14 (approximately half of which from MSs and the other half from SBHs), i.e. the IMBH does not grow by a significant amount, suggesting that the IMBH mass growth has little or no impact on the event rates.


\subsection{Dependence on physical parameters}\label{sec:dependence_on_parameters}

In order to assess the dependence of the results on the physical parameters of our system, we ran a suite of control simulations in which we varied several quantities. In particular, we explored the effects of varying (i) the initial stellar density profile (Section~\ref{sec:change_in_the_density_profile}) and (ii) the initial mass of the central IMBH (Section~\ref{sec:imbh_mass_change_variat}). In Section~\ref{sec:other_changes}, we further appraised the impact of having a different mass (and, in the case of MSs and WDs, corresponding radius) of the stellar-mass objects, a different size of the capture radius for GW events (i.e. varying $\alpha$), a different absorption process for WDs (EMRIs instead of TDEs), and a different stellar population age.


\subsubsection{The initial stellar density profile}\label{sec:change_in_the_density_profile}

The initial stellar density profile likely plays an important role in the determination of the absorption event rates. In this section, we study the evolution of the system when we replace the \citeauthor{Jaffe_1983} model ($\gamma=2$ in equation~\ref{eq:dehnen_model}) with a \citeauthor{Hernquist_1990} model ($\gamma=1$), keeping the initial cluster mass and half-mass radius equal to $10^8$~M$_{\sun}$ and 100~pc, respectively, thus setting up a model with an initial scale radius $a = 41.4$~pc (as opposed to 100~pc).

The average absorption event rates in this alternative scenario are listed in Table~\ref{tab:event_rates_init_conf_mf_to_nf}, whereas Fig.~\ref{fig:event_rate_init_conf_mf_to_nf} shows the rate of events as a function of time. A \citeauthor{BahcallWolf1976} cusp develops within the IMBH influence radius ($\approx$2 pc at the beginning of the integration, $\approx$2.3 pc at the end), as expected (Fig.~\ref{fig:density_1gyr_mf_to_nf}). The event rate for each species drops by a factor $\sim$20--30 with respect to the default case. This can be explained in terms of the different density at small scales (compare the two top panels in Fig.~\ref{fig:density_1gyr_mf_to_nf}): as an example, the average initial (final) total density within a radius of 1 pc is $2.4 \times 10^5$~M$_{\sun}$~pc$^{-3}$ ($7.9 \times 10^4$~M$_{\sun}$~pc$^{-3}$) for the \citeauthor{Jaffe_1983} case, and $1.4 \times 10^4$~M$_{\sun}$~pc$^{-3}$ ($7.8 \times 10^3$~M$_{\sun}$~pc$^{-3}$) in the \citeauthor{Hernquist_1990} case, i.e. a factor of $\sim$20 ($\sim$10) smaller; the proportional dependence of the event rates on the inner density is expected from the loss-cone theory \citep[e.g.][]{Merritt_2013}. If the original \citeauthor{Hernquist_1990} density profile had a larger central density in the beginning, it is reasonable to expect that the event rates would converge to the values obtained in the default scenario.

Note that, in this case, the event rates do not always decrease with time: there is typically an initial rise that culminates at $\approx$0.2~Gyr, followed by a relatively slow decay. This behaviour is a result of a sequence of events: first, the cusp develops within the IMBH influence radius, resulting in the growth of the event rates (in this scenario, the initial profile is less steep than the \citeauthor{BahcallWolf1976} cusp, as opposed to the default case); after it is fully developed, the whole system again slowly expands in a self-similar fashion, thus the central density drops, and so does the event rate of all species. It is interesting that the late-time evolution of the event rates is similar to that of the default scenario, with just a different normalization, owing to the difference in the central mass density of the two systems.

Here, the IMBH only grows by $\approx$700~$\msun{}$ by the end of the integration. Most of the growth is due to accretion of MSs and SBHs (in almost equal part), as in the default case.


\subsubsection{The initial mass of the intermediate-mass black hole}\label{sec:imbh_mass_change_variat}

\begin{figure}
\centering
\includegraphics[width=0.45\textwidth]{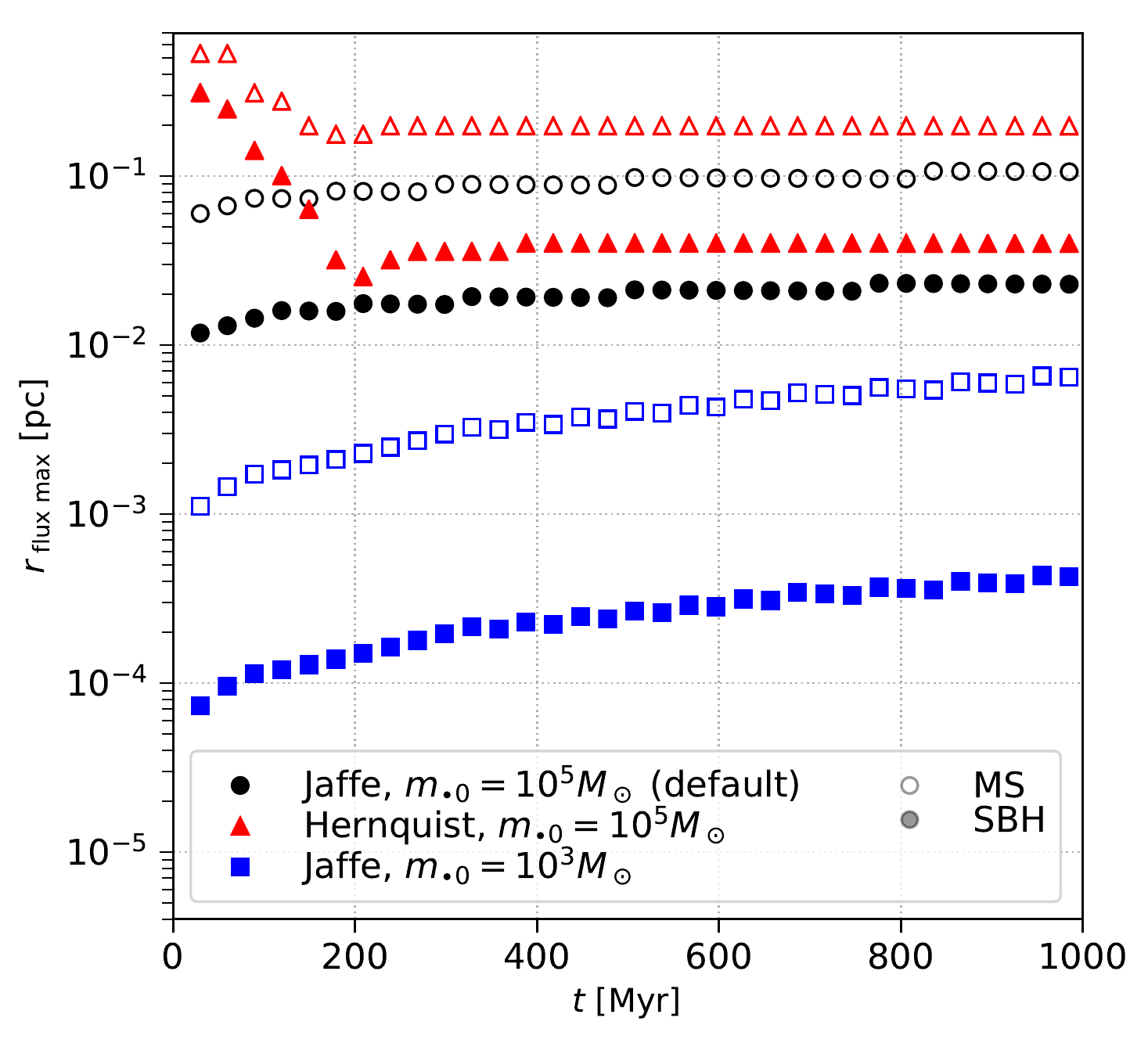}
\caption{Radius at which the flux of events peaks, for three different scenarios. Such radius depends on the considered species, and in particular, here we show its value for MSs (empty symbols) and SBHs (full symbols). Most of the flux comes from a radius that is roughly one order of magnitude smaller than the IMBH influence radius (evaluated at late times) for MSs, and from a radius which is nearly 5 (15) times smaller than the MS maximum flux radius if we consider SBHs, in the cases for which the IMBH has $m_{\bullet0} = 10^5\,\msun$ ($10^3\,\msun$).}
\label{fig:r_max_flux}
\end{figure}

Given the uncertainty on the mass of the IMBH hosted in the system, we here explore the impact on the event rates when assuming that the IMBH has an initial mass $m_{\bullet0} = 10^3$~M$_{\sun}$, i.e. 100 times smaller than in the default scenario. The  IMBH mass is a critical quantity for the estimation of the TDE and GW event rates, as both the tidal and capture radii grow with the IMBH mass, even if in a different fashion (see equations~\ref{eq:tidal_radius} and \ref{eq:capt_radius_compact_3}).  Table~\ref{tab:event_rates_init_conf_mf_to_nf} and Fig.~\ref{fig:event_rate_init_conf_mf_to_nf} report the rate of events in this configuration, showing that lowering the IMBH mass brings to a drop -- compared to the default scenario -- in the event rates for all species, although by vastly different amounts. The drop is by a factor of $\sim$30 for MSs and WDs, the tidally disrupted objects, and by a factor of $\sim$90 for NSs. On the contrary, GW events associated with SBHs drop only by a factor of $\sim$3. In addition, in this situation the IMBH grows by a substantial amount, enhancing its mass by a factor of $\sim$4 in 1~Gyr. Nearly 91 per cent of such growth is to be attributed to SBHs, whereas MSs grow the IMBH mass by about 8 per cent.

The crucial contribution of SBHs is likely due to the following facts: (i) they start to dominate the density profile at larger radii (Fig.~\ref{fig:density_1gyr_mf_to_nf}) and (ii) the radius from which most of the SBH flux comes from, shown in Fig.~\ref{fig:r_max_flux}, is much smaller in this situation (almost by a factor of 100), compared to the other two discussed scenarios. It follows that SBHs contribute more to the total capture rate.

We also note that the IMBH influence radius increases from an initial value of $\sim$$2 \times $10$^{-3}$~pc to $\sim$0.12~pc by the end of the integration, in response to the mass growth of the IMBH itself and to the evolution of the inner density profile.


\subsubsection{Robustness of the results}\label{sec:other_changes}

In the previous sections, we computed the absorption rates of different types of stellar remnants after fixing the values of several quantities related to both the stellar-mass objects and the IMBH. Here, we show that our choice of parameters does not significantly affect the results.

We first varied the mass of MSs (between 0.4 and 0.8~M$_{\sun}$), WDs (between 0.7 and 1.3~M$_{\sun}$), NSs (between 1.6 and 3~M$_{\sun}$), and SBHs (between 10 and 30~M$_{\sun}$). When varying the mass of MSs and WDs, we altered the stellar radius accordingly (see Table~\ref{tab:evolved_kroupa_mf_to_nf}). To better assess the effect of the change in each stellar population, we modified the stellar masses in a separate run for each component.\footnote{In order to keep fixed both the number fractions and the total number of objects, the mass of the stellar cluster has to vary, although by a little amount ($\sim$30 per cent when varying the MS mass, $\sim${1} per cent in the WD case, and $\lesssim$1 per cent in the NS and SBH cases).}

Given the uncertainty in the size of the capture radius (which depends on quantities that we cannot follow, such as the orbital properties of the inspiralling body and the IMBH spin), we then assessed the dependence of our results on the choice of the parameter $\alpha$ (see equation~\ref{eq:capt_radius_compact_3}), using the values listed in Table~\ref{tab:change_radii}.

\begin{table}
\centering
\begin{tabular}{cc}
\hline
Particle (Event) & Absorption radius ($\mathrm{pc}$) \T \B\\
\hline
MS (TDE) & $7.4\times 10^{-7}$ \T \B\\
NS, SBH (GW), $\alpha=2$ & $9.6 \times 10^{-9}$ \T \B\\
NS, SBH (GW), $\alpha=6$ & $2.9 \times 10^{-8}$ \T \B\\
NS, SBH (GW), $\alpha=8$ & $3.8 \times 10^{-8}$ \T \B\\
NS, SBH (GW), $\alpha=11.6$ & $5.6 \times 10^{-8}$ \T \B\\
WD (TDE) & $8.3 \times 10^{-9}$ \T \B\\
WD (GW), $\alpha = 8$ & $3.8 \times 10^{-8}$ \T \B\\
\hline
\end{tabular}
\caption{Initial absorption radii for the different types of events and parameters choices, for an IMBH with $m_{\bullet 0} = 10^5$~M$_{\sun}$. The values are obtained from equations~\eqref{eq:tidal_radius} and \eqref{eq:capt_radius_compact_3}, where for MSs and WDs the default values listed in Table~\ref{tab:evolved_kroupa_mf_to_nf} were used.}
\label{tab:change_radii}
\end{table}

\begin{table*}
\centering
\begin{tabular}{ccccccccccccccc}
\multicolumn{2}{c}{} & \multicolumn{2}{c}{\shortstack{$m_{\rm MS}$\\ ($\mathrm{M}_{\sun}$)}} & \multicolumn{2}{c}{\shortstack{$m_{\rm WD}$\\ ($\mathrm{M}_{\sun}$)}} & \multicolumn{2}{c}{\shortstack{$m_{\rm NS}$\\ ($\mathrm{M}_{\sun}$)}} & \multicolumn{2}{c}{\shortstack{$m_{\rm SBH}$\\ ($\mathrm{M}_{\sun}$)}} & \multicolumn{3}{c}{\shortstack{Parameter $\alpha$ \\ of equation~\eqref{eq:capt_radius_compact_3}}} & \shortstack{WD\\ absorption process} & \shortstack{Age\\ (Gyr)} \T \B \\
\cline{3-15}
\multicolumn{2}{c}{} & 0.4 & 0.8 & 0.7 & 1.3 & 1.6 & 3.0 & 10 & 30 & 2 & 6 & 11.6 & EMRI & {5} \T \B \\
\hline
\parbox[t]{2mm}{\multirow{4}{*}{\rotatebox[origin=c]{90}{per cent diff.}}} & MS & -25.2 & 28.6 & 0.0 & -0.1 & 0.2 & -0.4 & 42.6 & -22.2 & -2.7 & -0.6 & 1.0 & 0.0 & -4.1 \T \B \\
& WD  & -3.9 & 2.1 & 18.7 & -26.4 & 0.3 & -0.5 & 73.3 & -27.1 & -3.0 & -0.7 & 1.1 & 129.5 & 163.7 \T \B \\
& NS & -12.7 & 9.9 & -0.1 & -0.1 & -22.2 & 27.2 & 152.1 & -37.9 & -54.2 & -14.5 & 25.5 & 0.0 & -1.5 \T \B \\
& SBH & -30.3 & 28.3 & -0.7 & 0.6 & -0.3 & 0.2 & 14.5 & -14.5 & -45.7 & -11.3 & 18.6 & -0.1 & -0.5 \T \B \\
\hline
\end{tabular}
\caption{Percentage differences in the total number of events at 1~Gyr, with respect to the default case (whose numbers are listed in the third column of Table~\ref{tab:event_rates_init_conf_mf_to_nf}), when changing stellar quantities, absorption recipes, and stellar population age. The parameters of the default model are: $m_{\rm MS} = 0.6$~M$_{\sun}$, $m_{\rm WD} = 1.0$~M$_{\sun}$, $m_{\rm NS} = 2.3$~M$_{\sun}$, $m_{\rm SBH} = 20$~M$_{\sun}$, $\alpha = 8$, the absorption process undergone by WDs is a TDE, and the stellar population age is 1~Gyr.}
\label{tab:variation_percentages}
\end{table*}

We also investigated the impact of having a different absorption process for WDs, by assuming that, instead of undergoing a TDE (the default case), they produce an EMRI (in practice, employing equation~\ref{eq:capt_radius_compact_3} instead of equation~\ref{eq:tidal_radius}).

Finally, we computed how our results would change if we considered an older stellar population, by assuming an age of 5~Gyr (instead of the default value of 1~Gyr), i.e. considering a cluster turnoff mass of 1.320~M$_{\sun}$.

In Table~\ref{tab:variation_percentages}, we report the relative variation in the total event rates with respect to the default scenario, for all these cases. The event rates as a function of time (not shown) vary by similar amounts.

Altering the mass of WDs and NSs is inconsequential, except for a very mild increase (decrease) of $\sim$20--30 per cent in the event rate of the NSs when we increase (decrease) the NS particle mass, and a comparable decrease (increase) of $\sim$20--30 per cent in the event rate of the WDs when we increase (decrease) the WD particle mass.
Increasing (decreasing) the mass of MSs slightly enhances (reduces) the  event rates for all species, with changes varying from a few per cent in the case of WDs, to $\sim$10~per cent for NSs, to $\sim$30~per cent for MSs and SBHs. Varying the mass of SBHs has the largest effect of all, with positive (negative) changes in the event rates -- ranging between $\sim$15 per cent for SBHs and $\sim$150 per cent for NSs -- when the particle mass decreases (increases).

Varying the definition of the capture radius has the expected result of reducing (enhancing) the event rates when decreasing (increasing) $\alpha$. We note, however, that the change is at most $\sim$55~per cent.

Changing the absorption process for WDs has virtually no effect on the other species and only a mild increase in the event rates ($\sim$130~per cent), due to the fact that the absorption radius for TDEs is smaller than that of EMRIs (see Table~\ref{tab:change_radii}).

Increasing the stellar population age from 1 to 5~Gyr has almost no effect on the event rates, except for the TDEs from WDs, which increase by $\sim$160 per cent, due to the fact that an older stellar population contains a larger number of WDs (in this comparison, four times larger in number fraction). Clumpy galaxies at $z<1$ are rare, therefore 5 Gyr is the upper limit on the stellar population age. This implies that, if we consider only a change in stellar population age, the increase of TDEs from WDs just reported is the maximum possible variation.

Note that the variation in the event rates discussed here is difficult to interpret, owing to the fact that each variation impacts (i) the relaxation rate, (ii) the degree of mass segregation, and (iii) the structure of the models. For this reason, we do not attempt here to enter in detail on this aspect.

What we can clearly see is that, overall, absorption event rates are only mildly sensible to variations in the mass of the stellar-mass objects, to the definition of the capture radius, to the type of absorption process of WDs, and to the assumed stellar population age, as they change at most by a factor of $\lesssim$3.


\section{Discussion}\label{sec:discussion}

In this paper, we studied the rates of TDEs and GW events about an IMBH that could inhabit the centre of a massive, star-forming clump as the ones that are typically both observed and predicted by simulations at $z\lesssim 3$. However, we did not address how the IMBH could have formed or found itself within this system. Although many possible scenarios are possible (see Section~\ref{sec:introduction}), here we discuss the most favoured.

A particularly interesting scenario for the formation of IMBHs in the explored systems is via runaway collisions \citep[e.g.][]{Portegies-Zwart2004}: in particular, \citet{Portegies-Zwart2002} showed that, if the system has a half-mass relaxation time smaller than $\sim$25~Myr (i.e. roughly the lifespan of the most massive stars in the cluster), mass segregation could induce stellar collisions that can produce a star whose mass can reach $\sim$10$^{-3}$ cluster masses. This supermassive star is then expected to turn into an IMBH. Note that, in our system, runaway collisions could not occur on the scale of the entire cluster, whose half-mass relaxation time is longer than a Hubble time, but rather in the innermost region ($\sim$0.1~pc), where the relaxation time-scale is $\lesssim$25~Myr in the assumption of an initial \citeauthor{Jaffe_1983} profile.

An alternative possibility is that an IMBH is brought in the clumpy host galaxy via a minor merger; if the main galaxy hosts a large number of massive clumps, as in \citet{Tamburello_et_al_2015}, one of them could capture the wandering IMBH which would then sink in the cluster centre via dynamical friction. Several close interactions between a secondary BH and massive clumps are indeed observed in \citet{Tamburello_et_al_2017}, wherein a central and a secondary BH were added to the galaxies of \citet{Tamburello_et_al_2015} to study BH pairing time-scales.

\subsection{Fraction of direct plunges}\label{sec:plunges}

In Section~\ref{sec:results}, we reported the rate of gravitational captures due to the central IMBH. However, when a compact object enters the IMBH ISCO, it can do so either almost radially (i.e. via a so-called direct plunge), or via the gradual inspiral induced by GWs (via an EMRI or IMRI). Only in the latter case, the event could be detected by future facilities such as the forthcoming LISA.  Although a careful analysis of the EMRI/IMRI and direct-plunge orbits is beyond the scope of this paper, and the GW inspiral is not followed in the current framework, here we propose a simple analysis to obtain at least a rough estimate of the fraction of plunges. 

Following \citet{Amaro-Seoane2018} and \citet{Bortolas2019}, we can estimate the critical semimajor axis $a_{\rm p}$ above which only plunges can occur as

\begin{equation}
    a_{\rm p} = 0.18 \ {\rm pc}\ %
    C^{\sfrac{2}{3}} \left(\frac{t_{\rm rel}}{\rm 1\ Gyr}\right)^{\sfrac{2}{3}}%
    \left(\frac{m_\star}{20 \msun{}}\right)^{\sfrac{2}{3}}%
    \left(\frac{m_\bullet}{10^5\msun{}}\right)^{\sfrac{-1}{3}},
\end{equation}

\noindent where $C = 0.5$ \citep[][]{Merritt_et_al_2011,Brem_et_al_2014} is a constant that defines how much shorter the GW emission time-scale should be, compared to the two-body relaxation time, in order for a potential EMRI/IMRI not to be deflected by two-body relaxation; $t_{\rm rel}$ is the relaxation time-scale at $a_{\rm p}$; and $m_\star$ is the mass of the objects that drive two-body relaxation (i.e. SBHs at the small scales considered for this process). In order to obtain $a_{\rm p}$, we estimate $ t_{\rm rel}(E)\sim \mathcal{D}^{-1}(E)$, where $\mathcal{D}(E)$ is the sum of the angular momentum diffusion coefficients associated with each species in the Fokker--Planck treatment; $\mathcal{D}^{-1}(E)$ is evaluated at $a_{\rm p}$ by recursive iteration. Finally, we compute the fraction of direct plunges as a function of time as the fraction of GW events associated with SBHs or NSs coming from a semimajor axis larger than $a_{\rm p}$ (more precisely, an energy $E$ larger than the energy associated with $a_{\rm p}$). 

\begin{figure}
\centering
\includegraphics[width=0.42\textwidth]{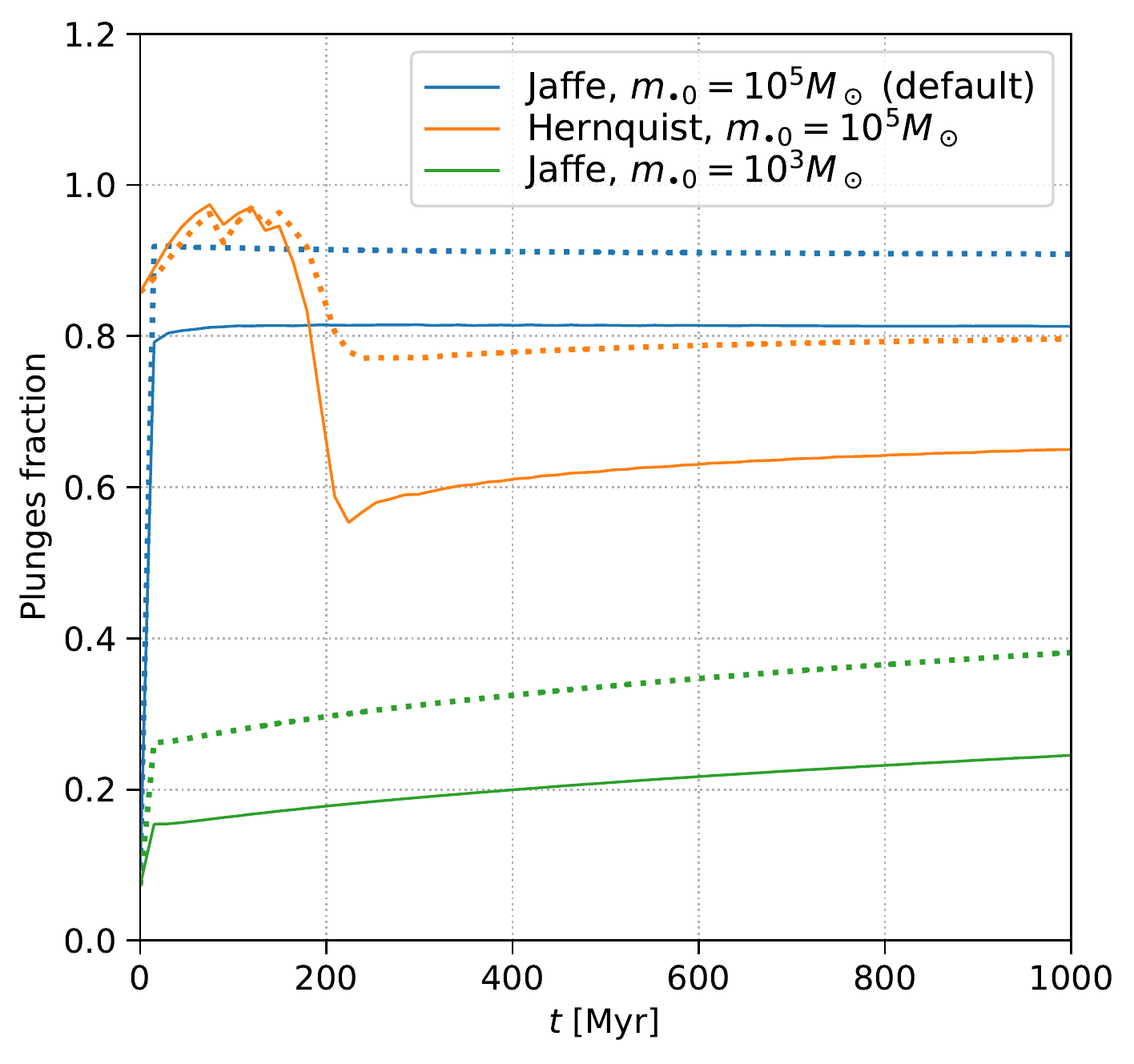}
\caption{Fraction of plunges with respect to the total number of GW events as a function of time for NSs (dotted) and SBHs (solid), for the three main scenarios discussed in this work.}
\label{fig:plunges_fractions}
\end{figure}

The fraction of plunges as a function of time for the three main scenarios explored in this work is shown in Fig.~\ref{fig:plunges_fractions}. The fraction of plunges is in the range of $\sim$0.6--0.9 in the two cases with $m_{\bullet0} = 10^5$~M$_{\sun}$, whereas it remains $\lesssim$0.4 if the central IMBH has an initial mass of $10^3\,\msun{}$. In all cases, the plunge fraction for NSs is larger than that for SBHs, as the lower mass associated with NSs renders their inspiral time-scale longer, making them more susceptible to deflections by two-body relaxation.

\subsection{Cosmological event rates}

Clumpy galaxies represent a significant fraction of the population of UV-bright star-forming galaxies, and their abundance peaks at the same epoch at which the cosmic star formation history reaches its maximum, at around $z \sim 2$ \citep{Shibuya_et_al_2016}. We set out to compute the expected cosmological IMRI rate resulting from SBHs accreting on to a $10^5$~M$_{\sun}$ IMBH at the centre of star-forming clumps, assuming that this is present in at least the most massive clumps ($\sim$10$^8$~M$_{\sun}$). Clumpy galaxies such as those simulated by \citet{Tamburello_et_al_2015} and \citet{Tamburello_et_al_2017} contain 1--5 of such massive clumps, which is actually a conservative result compared to other published simulations \citep[see e.g.][]{Mandelker_et_al_2014}. Conservatively, we will assume just one such massive clump per clumpy galaxy. To compute the cosmological IMRI rate, we count events in the (comoving) cosmological volume between $z=1$ and 3. This is because at lower and higher redshift the number of clumpy galaxies drops drastically, whereas it is estimated to be between 40 and 60 per cent of the star-forming galaxy population within such redshift range \citep{Shibuya_et_al_2016}. Note that clumpy galaxies are present also at lower and higher redshift, hence our estimate will be conservative also on this side. We note, however, that at $z > 3$ IMRIs occurring on an IMBH of the mass considered here would be hardly detectable by LISA due to low signal-to-noise ratio \citep[][]{Babak2017}. With the chosen volume, our estimate should thus yield a lower limit on the number of such IMRIs that should be detectable by LISA.

In order to compute that, we take into account that, when the IMBH mass is $10^5$~M$_{\sun}$, $\sim$30 per cent of such events are expected to be accessible to the LISA observatory, the exact number depending on a range of assumptions (see \citealt{Babak2017}, and in particular their fig.~10, for more details on this). Moreover, as discussed in Section~\ref{sec:plunges}, we need to subtract the direct plunges which, in the default scenario, account for $\sim$80 per cent of the GW events.

The cosmological LISA-detectable IMRI event rate in clumpy galaxies is thus given by $R_{\rm IMRI, cosmo}= n_{\rm SF} f_{\rm clumpy} r_{\rm IMRI} V_{\Delta z} f_{\rm det}$, where $n_{\rm SF}$ is the number density of star-forming galaxies, which we assume to be constant in the chosen redshift range \citep[a reasonable approximation within a factor of 2; see, e.g.][]{Brammer_et_al_2011}, $f_{\rm clumpy}$ is the fraction of clumpy galaxies (which we also assume to be constant: $f_{\rm clumpy} = 0.5$), $V_{\Delta z}$ is the comoving cosmological volume between $z=3$ and 1 \citep[$\sim$1000~Gpc$^3$;][]{Wright_2006,Planck_et_al_2018}, $r_{\rm IMRI}$ is the IMRI rate per clump computed in the previous sections ($4 \times 10^{-7}$~yr$^{-1}$, see Fig.~\ref{fig:event_rate_init_conf_mf_to_nf}), and $f_{\rm det}$ is a suppression factor, which we estimate to be roughly 0.05, that accounts for (i) the detectability of such events by LISA and (ii) the fact that some of these events would be direct plunges. Note that, according to \citet{Brammer_et_al_2011}, the number density of star-forming galaxies varies slightly as a function of luminosity/stellar mass considered, but by less than a factor of 2 in the mass range in which a disc is massive enough to fragment \citep[4--$10 \times 10^{11}$~M$_{\sun}$;][]{Tamburello_et_al_2015}, being around $\sim$$2 \times 10^{-4}$~Mpc$^{-3}$. For the conservative assumption of one star-forming clump per galaxy that hosts a central IMBH, we obtain $R_{\rm IMRI, cosmo} \sim 2$ events per year.

A similar calculation can be performed for TDEs. Since our computations show that the TDE rate is $\sim$100 times higher than the IMRI rate, we expect an intrinsic rate $R_{\rm TDE, cosmo}$ of a few thousand events per year in the cosmological volume between $z = 1$ and 3. Even though current observations of TDEs are all within $z \lesssim 1$, future observatories should be able to detect several events also at high redshift. For example, SKA may be able to detect several hundred TDEs per year below $z \sim 2.5$ (with a peak rate redshift $\sim 0.5$). Using follow-up observations by X-ray facilities such as Athena, the number of identified TDEs could be $\sim$400 per year, almost up to $z = 2$ \citep[][]{Donnarumma_et_al_2015}.  

\subsection{Model limitations and caveats}

The approach adopted in this paper to address the rates of TDEs and GW events is necessarily approximated in many ways. Here, we discuss the main limitations and caveats. 

First of all, we opted  for a Fokker--Planck approach, which is very fast and computationally cheap. However, its advantages come at a price: in particular, the Fokker--Planck method adopted in {\fontfamily{qcr}\selectfont phaseflow} handles a 1D distribution function that only depends on energy, and not on the angular momentum; the dependence of the Fokker--Planck solution on angular momentum in the context of collisional stellar systems has been shown to be weak \citep{Cohn1980} and, for the IMBH masses we are addressing, the TDE rates have been shown not to get crucially suppressed by e.g. an initial tangential anisotropy in the velocity space \citep{Lezhnin2015}. More in general, {\fontfamily{qcr}\selectfont phaseflow} can only handle spherically symmetric systems with isotropic velocity distributions. Neglecting a possible net rotation, radial anisotropy in the velocities, and deviations from spherical symmetry likely results in an underestimate of the loss-cone flux \citep{Holley-Bockelmann2006,Holley-Bockelmann2015,Lezhnin2016,Stone2020}; in this sense, our estimates can be considered lower limits to the actual event rates. However, one should also consider that the regions near the IMBH are expected to turn into  spherical and isotropic owing to two-body relaxation  effects,  and possibly also to the orbital scattering of stars owing to the presence of the IMBH; both effects occur over a time-scale of the order of the relaxation time \citep{Lake1983,Gerhard1985,MerrittPoon2004,Bortolas2018}.

Furthermore, our treatment assumed an initial lack of mass segregation in the system (i.e. all components shared the same initial density profile); in fact, the central regions of our system relax very quickly in our simulations, and they only retain memory of the initial central density normalization, thus we do not expect this aspect to significantly impact our results. The adoption of four distinct stellar families instead of a smooth and non-discretized mass function is also an approximation, but it is already a significant improvement over the more common approximation of a single-mass family.

An additional caveat concerns the IMBH at the centre of the clump. First of all, we do not address the formation of the IMBH; the occupation fraction of massive clumps by IMBHs is very uncertain; however, our estimate of the event rates detailed in the previous section is conservative on the former aspect (we assumed only one clump hosting an IMBH per galaxy). Clearly, the dynamics of the IMBH cannot be captured in our idealized simulations: in fact, the IMBH Brownian wandering resulting from two-body interactions with the surrounding stars  \citep{Lin1980, Merritt2005} could influence the event rates in real physical systems, but it was not captured in the Fokker--Planck simulations. Brownian motion is expected to be significant in the system for which we assumed a low IMBH mass ($m_{\bullet0}=10^3 \,\msun{}$), while it has been shown to be negligible for central objects with masses $\sim$$10^5\,\msun$ in the context of massive BH binaries\footnote{Massive BH binaries experience superelastic scatterings with the surrounding stars, thus the Brownian wandering in \citet{Bortolas2016} is enhanced compared to single massive BHs \citep{Merritt2001}.} \citep{Bortolas2016}.

Connected to this, one should also account for the GW recoil that the IMBH could undergo in response to the anisotropic emission of gravitational radiation following an EMRI or IMRI. This phenomenon is well studied and known to scale with the spins and the mass ratio between the two merging compact objects. The magnitude of the kick associated with small mass ratios is typically very low \citep{Baker2008,Morawski2018}, much lower than the escape speed of the system considered here.

In our treatment, we could not account for the presence of gas, and we assumed that this component either collapses into stars or is evaporated from the clump by means of stellar feedback. This is likely a good approximation after the first supernovae in the system occurred, thus after a few tens of Myr from the beginning of our integration.

Finally, our treatment of gravity was purely Newtonian, and did not account for general relativistic effects. In addition, we did not take into account resonant relaxation \citep{Rauch1996}, but its effect has been shown to be virtually washed out by its coupling with relativistic precession \citep{Merritt2015, Alexander2017}, thus this process is not expected to significantly affect the event rates.

\section{Conclusion}\label{sec:conclusion}

In this paper, we explored the rates of TDEs and GW events that could occur about IMBHs hosted in massive star-forming clumps, typical structures that are found in many galaxies in the high redshift Universe. Our reference massive clump was modelled so that its proprieties are in good agreement with what found in simulations of clumpy galaxies at $z\approx 1$--3 \citep{Tamburello_et_al_2015, Tamburello_et_al_2017}. In fact, clumpy galaxies constitute an abundant fraction (about 50 per cent) of the star-forming galactic population in this range of redshifts \citep{Shibuya_et_al_2016}. 

We found that the TDE rate for MSs is in the range of $10^{-6}$--$10^{-5}$~yr$^{-1}$ per massive clump, depending on the mass of the central IMBH and on the density profile of the clump. We estimated the rate of GW events involving SBHs to be around $10^{-7}$~yr$^{-1}$ for IMBHs with masses of $\sim$10$^3$--$10^5$~M$_{\sun}$; such rate decreases by an order of magnitude if the initial density profile in the inner regions is assumed to scale as $r^{-1}$ instead of $r^{-2}$. We estimated a relevant fraction (at least $\approx$20 per cent for SBHs) of GW events to be EMRIs and IMRIs, thus to be in principle detectable by the forthcoming LISA mission.

Finally, we estimated that the forthcoming LISA observatory will be able to detect nearly 2 GW events per year originating from such star-forming clumps; we also estimated the intrinsic rate of TDEs of MSs in these clumps, obtaining a few $10^3$ events per year; a fraction of these can be observed by forthcoming instruments, and in particular by SKA in combination with Athena. 

We highlight that, in the future, the observed event rates in the redshift range $z\approx 1$--3 could be adopted to constrain the occurrence of massive clumps hosting IMBHs.

In conclusion, both TDE and EMRI/IMRI detections will be crucial tools to explore the occupation fraction of IMBHs in massive star-forming clumps in the early Universe, and will provide new details on the origin and evolution of massive BHs across the cosmic time.

\section*{Acknowledgements}
We warmly thank the reviewer for the very useful report that definitely improved the quality of our paper, and Eugene Vasiliev for help with the use of {\fontfamily{qcr}\selectfont phaseflow} and very useful comments and suggestions in the interpretation of the results. EB, PRC, and LM acknowledge support from the Swiss National Science Foundation under the grant 200020\_178949.

\section*{Data Availability Statement}
The data underlying this article will be shared on reasonable request to the corresponding author.

\scalefont{0.94}
\setlength{\bibhang}{1.6em}
\setlength\labelwidth{0.0em}
\bibliographystyle{mnras}
\bibliography{generation_of_gravitational_waves_and_tidal_disruptions_in_clumpy_galaxies}

\begin{thebibliography}{}
\makeatletter
\relax
\def\mn@urlcharsother{\let\do\@makeother \do\$\do\&\do\#\do\^\do\_\do\%\do\~}
\def\mn@doi{\begingroup\mn@urlcharsother \@ifnextchar [ {\mn@doi@}
  {\mn@doi@[]}}
\def\mn@doi@[#1]#2{\def\@tempa{#1}\ifx\@tempa\@empty \href
  {http://dx.doi.org/#2} {doi:#2}\else \href {http://dx.doi.org/#2} {#1}\fi
  \endgroup}
\def\mn@eprint#1#2{\mn@eprint@#1:#2::\@nil}
\def\mn@eprint@arXiv#1{\href {http://arxiv.org/abs/#1} {{\tt arXiv:#1}}}
\def\mn@eprint@dblp#1{\href {http://dblp.uni-trier.de/rec/bibtex/#1.xml}
  {dblp:#1}}
\def\mn@eprint@#1:#2:#3:#4\@nil{\def\@tempa {#1}\def\@tempb {#2}\def\@tempc
  {#3}\ifx \@tempc \@empty \let \@tempc \@tempb \let \@tempb \@tempa \fi \ifx
  \@tempb \@empty \def\@tempb {arXiv}\fi \@ifundefined
  {mn@eprint@\@tempb}{\@tempb:\@tempc}{\expandafter \expandafter \csname
  mn@eprint@\@tempb\endcsname \expandafter{\@tempc}}}

\bibitem[\protect\citeauthoryear{Abbott et~al.,}{Abbott
  et~al.}{2016}]{Abbott_2016}
Abbott B.~P.,  et~al., 2016, \mn@doi [Phys. Rev. Lett.]
  {10.1103/PhysRevLett.116.061102}, \href
  {http://adsabs.harvard.edu/abs/2016PhRvL.116f1102A} {116, 061102}

\bibitem[\protect\citeauthoryear{{Abbott} et~al.,}{{Abbott}
  et~al.}{2020}]{Abbott_2020}
{Abbott} R.,  et~al., 2020, \mn@doi [\prl] {10.1103/PhysRevLett.125.101102},
  \href {https://ui.adsabs.harvard.edu/abs/2020PhRvL.125j1102A} {125, 101102}

\bibitem[\protect\citeauthoryear{{Alexander}}{{Alexander}}{2017}]{Alexander2017}
{Alexander} T.,  2017, in Journal of Physics Conference Series. p. 012019
  (\mn@eprint {arXiv} {1702.00597}), \mn@doi{10.1088/1742-6596/840/1/012019}

\bibitem[\protect\citeauthoryear{Alexander \& Hopman}{Alexander \&
  Hopman}{2009}]{Alexander_2009}
Alexander T.,  Hopman C.,  2009, \mn@doi [ApJ] {10.1088/0004-637X/697/2/1861},
  \href {http://adsabs.harvard.edu/abs/2009ApJ...697.1861A} {697, 1861}

\bibitem[\protect\citeauthoryear{{Amaro-Seoane}}{{Amaro-Seoane}}{2018}]{Amaro-Seoane2018}
{Amaro-Seoane} P.,  2018, \mn@doi [Living Reviews in Relativity]
  {10.1007/s41114-018-0013-8}, \href
  {https://ui.adsabs.harvard.edu/abs/2018LRR....21....4A} {21, 4}

\bibitem[\protect\citeauthoryear{{Amaro-Seoane} \& {Preto}}{{Amaro-Seoane} \&
  {Preto}}{2011}]{Amaro-Seoane2011}
{Amaro-Seoane} P.,  {Preto} M.,  2011, \mn@doi [Classical and Quantum Gravity]
  {10.1088/0264-9381/28/9/094017}, \href
  {https://ui.adsabs.harvard.edu/abs/2011CQGra..28i4017A} {28, 094017}

\bibitem[\protect\citeauthoryear{{Amaro-Seoane}, Gair, Freitag, Miller, Mandel,
  Cutler  \& Babak}{{Amaro-Seoane} et~al.}{2007}]{Amaro-Seoane_2007}
{Amaro-Seoane} P.,  Gair J.~R.,  Freitag M.,  Miller M.~C.,  Mandel I.,  Cutler
  C.~J.,   Babak S.,  2007, \mn@doi [Class. Quantum Gravity]
  {10.1088/0264-9381/24/17/R01}, \href
  {http://adsabs.harvard.edu/abs/2007CQGra..24R.113A} {24, R113}

\bibitem[\protect\citeauthoryear{{Amaro-Seoane} et~al.,}{{Amaro-Seoane}
  et~al.}{2017}]{LISA_2017}
{Amaro-Seoane} P.,  et~al., 2017, arXiv e-prints, \href
  {https://ui.adsabs.harvard.edu/abs/2017arXiv170200786A} {p. arXiv:1702.00786}

\bibitem[\protect\citeauthoryear{Anninos, Fragile, Olivier, Hoffman, Mishra  \&
  Camarda}{Anninos et~al.}{2018}]{Anninos_2018}
Anninos P.,  Fragile P.~C.,  Olivier S.~S.,  Hoffman R.,  Mishra B.,   Camarda
  K.,  2018, \mn@doi [ApJ] {10.3847/1538-4357/aadad9}, \href
  {https://ui.adsabs.harvard.edu/abs/2018ApJ...865....3A/abstract} {865, 3}

\bibitem[\protect\citeauthoryear{{Babak} et~al.,}{{Babak}
  et~al.}{2017}]{Babak2017}
{Babak} S.,  et~al., 2017, \mn@doi [\prd] {10.1103/PhysRevD.95.103012}, \href
  {https://ui.adsabs.harvard.edu/abs/2017PhRvD..95j3012B} {95, 103012}

\bibitem[\protect\citeauthoryear{{Bahcall} \& {Wolf}}{{Bahcall} \&
  {Wolf}}{1976}]{BahcallWolf1976}
{Bahcall} J.~N.,  {Wolf} R.~A.,  1976, \mn@doi [\apj] {10.1086/154711}, \href
  {https://ui.adsabs.harvard.edu/abs/1976ApJ...209..214B} {209, 214}

\bibitem[\protect\citeauthoryear{{Baker}, {Boggs}, {Centrella}, {Kelly},
  {McWilliams}, {Miller}  \& {van Meter}}{{Baker} et~al.}{2008}]{Baker2008}
{Baker} J.~G.,  {Boggs} W.~D.,  {Centrella} J.,  {Kelly} B.~J.,  {McWilliams}
  S.~T.,  {Miller} M.~C.,   {van Meter} J.~R.,  2008, \mn@doi [\apjl]
  {10.1086/590927}, \href
  {https://ui.adsabs.harvard.edu/abs/2008ApJ...682L..29B} {682, L29}

\bibitem[\protect\citeauthoryear{{Baldassare}, {Reines}, {Gallo}  \&
  {Greene}}{{Baldassare} et~al.}{2015}]{Baldassare2015}
{Baldassare} V.~F.,  {Reines} A.~E.,  {Gallo} E.,   {Greene} J.~E.,  2015,
  \mn@doi [\apjl] {10.1088/2041-8205/809/1/L14}, \href
  {http://adsabs.harvard.edu/abs/2015ApJ...809L..14B} {809, L14}

\bibitem[\protect\citeauthoryear{{Baldassare}, {Dickey}, {Geha}  \&
  {Reines}}{{Baldassare} et~al.}{2020}]{Baldassare_et_al_2020}
{Baldassare} V.~F.,  {Dickey} C.,  {Geha} M.,   {Reines} A.~E.,  2020, \mn@doi
  [\apjl] {10.3847/2041-8213/aba0c1}, \href
  {https://ui.adsabs.harvard.edu/abs/2020ApJ...898L...3B} {898, L3}

\bibitem[\protect\citeauthoryear{{Barack} \& {Cutler}}{{Barack} \&
  {Cutler}}{2004}]{Barack_et_al_2004}
{Barack} L.,  {Cutler} C.,  2004, \mn@doi [\prd] {10.1103/PhysRevD.69.082005},
  \href {https://ui.adsabs.harvard.edu/abs/2004PhRvD..69h2005B} {69, 082005}

\bibitem[\protect\citeauthoryear{{Barack} et~al.,}{{Barack}
  et~al.}{2019}]{Barack_et_al_2019}
{Barack} L.,  et~al., 2019, \mn@doi [Classical and Quantum Gravity]
  {10.1088/1361-6382/ab0587}, \href
  {https://ui.adsabs.harvard.edu/abs/2019CQGra..36n3001B} {36, 143001}

\bibitem[\protect\citeauthoryear{{Barcons} et~al.,}{{Barcons}
  et~al.}{2012}]{Barcons2012}
{Barcons} X.,  et~al., 2012, arXiv e-prints, \href
  {https://ui.adsabs.harvard.edu/abs/2012arXiv1207.2745B/abstract} {p.
  arXiv:1207.2745}

\bibitem[\protect\citeauthoryear{{Bortolas} \& {Mapelli}}{{Bortolas} \&
  {Mapelli}}{2019}]{Bortolas2019}
{Bortolas} E.,  {Mapelli} M.,  2019, \mn@doi [\mnras] {10.1093/mnras/stz440},
  \href {https://ui.adsabs.harvard.edu/abs/2019MNRAS.485.2125B} {485, 2125}

\bibitem[\protect\citeauthoryear{{Bortolas}, {Gualandris}, {Dotti}, {Spera}  \&
  {Mapelli}}{{Bortolas} et~al.}{2016}]{Bortolas2016}
{Bortolas} E.,  {Gualandris} A.,  {Dotti} M.,  {Spera} M.,   {Mapelli} M.,
  2016, \mn@doi [\mnras] {10.1093/mnras/stw1372}, \href
  {https://ui.adsabs.harvard.edu/abs/2016MNRAS.461.1023B} {461, 1023}

\bibitem[\protect\citeauthoryear{{Bortolas}, {Gualandris}, {Dotti}  \&
  {Read}}{{Bortolas} et~al.}{2018}]{Bortolas2018}
{Bortolas} E.,  {Gualandris} A.,  {Dotti} M.,   {Read} J.~I.,  2018, \mn@doi
  [\mnras] {10.1093/mnras/sty775}, \href
  {https://ui.adsabs.harvard.edu/abs/2018MNRAS.477.2310B} {477, 2310}

\bibitem[\protect\citeauthoryear{{Brammer} et~al.,}{{Brammer}
  et~al.}{2011}]{Brammer_et_al_2011}
{Brammer} G.~B.,  et~al., 2011, \mn@doi [\apj] {10.1088/0004-637X/739/1/24},
  \href {https://ui.adsabs.harvard.edu/abs/2011ApJ...739...24B} {739, 24}

\bibitem[\protect\citeauthoryear{{Brem}, {Amaro-Seoane}  \& {Sopuerta}}{{Brem}
  et~al.}{2014}]{Brem_et_al_2014}
{Brem} P.,  {Amaro-Seoane} P.,   {Sopuerta} C.~F.,  2014, \mn@doi [\mnras]
  {10.1093/mnras/stt1948}, \href
  {https://ui.adsabs.harvard.edu/abs/2014MNRAS.437.1259B} {437, 1259}

\bibitem[\protect\citeauthoryear{{Carr}, {Kohri}, {Sendouda}  \&
  {Yokoyama}}{{Carr} et~al.}{2020}]{Carr_et_al_2020}
{Carr} B.,  {Kohri} K.,  {Sendouda} Y.,   {Yokoyama} J.,  2020, arXiv e-prints,
  \href {https://ui.adsabs.harvard.edu/abs/2020arXiv200212778C} {p.
  arXiv:2002.12778}

\bibitem[\protect\citeauthoryear{Casares \& Jonker}{Casares \&
  Jonker}{2014}]{Casares_2014}
Casares J.,  Jonker P.~G.,  2014, \mn@doi [Space Sci. Rev]
  {10.1007/s11214-013-0030-6}, \href
  {http://adsabs.harvard.edu/abs/2014SSRv..183..223C} {183, 223}

\bibitem[\protect\citeauthoryear{{Chandrasekhar}}{{Chandrasekhar}}{1939}]{Chandrasekhar_1939}
{Chandrasekhar} S.,  1939, {An Introduction to the Study of Stellar Structure}.
The University of Chicago Press

\bibitem[\protect\citeauthoryear{Chen, Girardi, Bressan, Marigo, Barbieri  \&
  Kong}{Chen et~al.}{2014}]{Chen_2014}
Chen J.-H.,  Girardi L.,  Bressan A.,  Marigo P.,  Barbieri M.,   Kong X.,
  2014, \mn@doi [MNRAS] {10.1093/mnras/stu1605}, \href
  {http://adsabs.harvard.edu/abs/2014MNRAS.444.2525C} {444, 2525}

\bibitem[\protect\citeauthoryear{{Cohn}}{{Cohn}}{1980}]{Cohn1980}
{Cohn} H.,  1980, \mn@doi [\apj] {10.1086/158511}, \href
  {https://ui.adsabs.harvard.edu/abs/1980ApJ...242..765C} {242, 765}

\bibitem[\protect\citeauthoryear{Dehnen}{Dehnen}{1993}]{Dehnen_1993}
Dehnen W.,  1993, \mn@doi [MNRAS] {10.1093/mnras/265.1.250}, \href
  {http://adsabs.harvard.edu/abs/1993MNRAS.265..250D} {265, 250}

\bibitem[\protect\citeauthoryear{{Dewdney}, {Hall}, {Schilizzi}  \&
  {Lazio}}{{Dewdney} et~al.}{2009}]{Dewdney_et_al_2009}
{Dewdney} P.~E.,  {Hall} P.~J.,  {Schilizzi} R.~T.,   {Lazio} T.~J.~L.~W.,
  2009, \mn@doi [IEEE Proceedings] {10.1109/JPROC.2009.2021005}, \href
  {https://ui.adsabs.harvard.edu/abs/2009IEEEP..97.1482D} {97, 1482}

\bibitem[\protect\citeauthoryear{{Donnarumma} \& {Rossi}}{{Donnarumma} \&
  {Rossi}}{2015}]{Donnarumma_et_al_2015}
{Donnarumma} I.,  {Rossi} E.~M.,  2015, \mn@doi [\apj]
  {10.1088/0004-637X/803/1/36}, \href
  {https://ui.adsabs.harvard.edu/abs/2015ApJ...803...36D} {803, 36}

\bibitem[\protect\citeauthoryear{Elmegreen, Elmegreen, Fernandez  \&
  Lemonias}{Elmegreen et~al.}{2009}]{Elmegreen_2009}
Elmegreen B.~G.,  Elmegreen D.~M.,  Fernandez M.~X.,   Lemonias J.~J.,  2009,
  \mn@doi [ApJ] {10.1088/0004-637X/692/1/12}, \href
  {http://adsabs.harvard.edu/abs/2009ApJ...692...12E} {692, 12}

\bibitem[\protect\citeauthoryear{{Emami} \& {Loeb}}{{Emami} \&
  {Loeb}}{2020}]{Emami_Loeb_2020}
{Emami} R.,  {Loeb} A.,  2020, \mn@doi [\jcap] {10.1088/1475-7516/2020/02/021},
  \href {https://ui.adsabs.harvard.edu/abs/2020JCAP...02..021E} {2020, 021}

\bibitem[\protect\citeauthoryear{{Event Horizon Telescope
  Collaboration}}{{Event Horizon Telescope
  Collaboration}}{2019}]{Event_Horizon_Telescope_Collaboration_2019}
{Event Horizon Telescope Collaboration} 2019, \mn@doi [ApJ]
  {10.3847/2041-8213/ab0ec7}, \href
  {https://ui.adsabs.harvard.edu/abs/2019ApJ...875L...1E/abstract} {875, L1}

\bibitem[\protect\citeauthoryear{{Fragione}, {Ginsburg}  \&
  {Kocsis}}{{Fragione} et~al.}{2018a}]{Fragione_Ginsburg_et_al_2018}
{Fragione} G.,  {Ginsburg} I.,   {Kocsis} B.,  2018a, \mn@doi [\apj]
  {10.3847/1538-4357/aab368}, \href
  {https://ui.adsabs.harvard.edu/abs/2018ApJ...856...92F} {856, 92}

\bibitem[\protect\citeauthoryear{{Fragione}, {Leigh}, {Ginsburg}  \&
  {Kocsis}}{{Fragione} et~al.}{2018b}]{Fragione_Leigh_et_al_2018}
{Fragione} G.,  {Leigh} N. W.~C.,  {Ginsburg} I.,   {Kocsis} B.,  2018b,
  \mn@doi [\apj] {10.3847/1538-4357/aae486}, \href
  {https://ui.adsabs.harvard.edu/abs/2018ApJ...867..119F} {867, 119}

\bibitem[\protect\citeauthoryear{Freitag, G{\"u}rkan  \& Rasio}{Freitag
  et~al.}{2006}]{Freitag_2006}
Freitag M.,  G{\"u}rkan M.~A.,   Rasio F.~A.,  2006, \mn@doi [MNRAS]
  {10.1111/j.1365-2966.2006.10096.x}, \href
  {http://adsabs.harvard.edu/abs/2006MNRAS.368..141F} {368, 141}

\bibitem[\protect\citeauthoryear{Gebhardt, Rich  \& Ho}{Gebhardt
  et~al.}{2005}]{Gebhardt_2005}
Gebhardt K.,  Rich R.~M.,   Ho L.~C.,  2005, \mn@doi [ApJ] {10.1086/497023},
  \href {http://adsabs.harvard.edu/abs/2005ApJ...634.1093G} {634, 1093}

\bibitem[\protect\citeauthoryear{{Generozov}, {Stone}, {Metzger}  \&
  {Ostriker}}{{Generozov} et~al.}{2018}]{Generozov_et_al_2018}
{Generozov} A.,  {Stone} N.~C.,  {Metzger} B.~D.,   {Ostriker} J.~P.,  2018,
  \mn@doi [\mnras] {10.1093/mnras/sty1262}, \href
  {https://ui.adsabs.harvard.edu/abs/2018MNRAS.478.4030G} {478, 4030}

\bibitem[\protect\citeauthoryear{{Gerhard} \& {Binney}}{{Gerhard} \&
  {Binney}}{1985}]{Gerhard1985}
{Gerhard} O.~E.,  {Binney} J.,  1985, \mn@doi [\mnras]
  {10.1093/mnras/216.2.467}, \href
  {https://ui.adsabs.harvard.edu/abs/1985MNRAS.216..467G} {216, 467}

\bibitem[\protect\citeauthoryear{Gerssen, {van der Marel}, Gebhardt,
  Guhathakurta, Peterson  \& Pryor}{Gerssen et~al.}{2002}]{Gerssen_2002}
Gerssen J.,  {van der Marel} R.~P.,  Gebhardt K.,  Guhathakurta P.,  Peterson
  R.~C.,   Pryor C.,  2002, \mn@doi [ApJ] {10.1086/344584}, \href
  {http://adsabs.harvard.edu/abs/2002AJ....124.3270G} {124, 3270}

\bibitem[\protect\citeauthoryear{Giersz, Leigh, Hypki, L{\"u}tzgendorf  \&
  Askar}{Giersz et~al.}{2015}]{Giersz_2015}
Giersz M.,  Leigh N.,  Hypki A.,  L{\"u}tzgendorf N.,   Askar A.,  2015,
  \mn@doi [MNRAS] {10.1093/mnras/stv2162}, \href
  {http://adsabs.harvard.edu/abs/2015MNRAS.454.3150G} {454, 3150}

\bibitem[\protect\citeauthoryear{Giesers et~al.,}{Giesers
  et~al.}{2018}]{Giesers_2017}
Giesers B.,  et~al., 2018, \mn@doi [MNRAS] {10.1093/mnrasl/slx203}, \href
  {https://ui.adsabs.harvard.edu/abs/2018MNRAS.475L..15G/abstract} {475, L15}

\bibitem[\protect\citeauthoryear{Gillessen, Eisenhauer, Trippe, Alexander,
  Genzel, Martins  \& Ott}{Gillessen et~al.}{2009}]{Gillessen_2009}
Gillessen S.,  Eisenhauer F.,  Trippe S.,  Alexander T.,  Genzel R.,  Martins
  F.,   Ott T.,  2009, \mn@doi [ApJ] {10.1088/0004-637X/692/2/1075}, \href
  {https://ui.adsabs.harvard.edu/abs/2009ApJ...692.1075G/abstract} {692, 1075}

\bibitem[\protect\citeauthoryear{G{\"u}ltekin, Miller  \&
  Hamilton}{G{\"u}ltekin et~al.}{2004}]{Gultekin_2004}
G{\"u}ltekin K.,  Miller M.~C.,   Hamilton D.~P.,  2004, \mn@doi [ApJ]
  {10.1086/424809}, \href {http://adsabs.harvard.edu/abs/2004ApJ...616..221G}
  {616, 221}

\bibitem[\protect\citeauthoryear{G{\"u}rkan, Freitag  \& Rasio}{G{\"u}rkan
  et~al.}{2004}]{Gurkan_fr_2004}
G{\"u}rkan M.~A.,  Freitag M.,   Rasio F.~A.,  2004, \mn@doi [ApJ]
  {10.1086/381968}, \href {http://adsabs.harvard.edu/abs/2004ApJ...604..632G}
  {604, 632}

\bibitem[\protect\citeauthoryear{G{\"u}rkan, Fregeau  \& Rasio}{G{\"u}rkan
  et~al.}{2006}]{Gurkan_2006}
G{\"u}rkan M.~A.,  Fregeau J.~M.,   Rasio F.~A.,  2006, \mn@doi [ApJ]
  {10.1086/503295}, \href {http://adsabs.harvard.edu/abs/2006ApJ...640L..39G}
  {640, L39}

\bibitem[\protect\citeauthoryear{Haas, Shcherbakov, Bode  \& Laguna}{Haas
  et~al.}{2012}]{Haas_2012}
Haas R.,  Shcherbakov R.~V.,  Bode T.,   Laguna P.,  2012, \mn@doi [ApJ]
  {10.1088/0004-637X/749/2/117}, \href
  {https://ui.adsabs.harvard.edu/abs/2012ApJ...749..117H/abstract} {749, 1385}

\bibitem[\protect\citeauthoryear{Haggard, Cool, Heinke, {van der Marel}, Cohn,
  Lugger  \& Anderson}{Haggard et~al.}{2013}]{Haggard_2013}
Haggard D.,  Cool A.~M.,  Heinke C.~O.,  {van der Marel} R.,  Cohn H.~N.,
  Lugger P.,   Anderson J.,  2013, \mn@doi [ApJ] {10.1088/2041-8205/773/2/L31},
  \href {http://adsabs.harvard.edu/abs/2013ApJ...773L..31H} {773, L31}

\bibitem[\protect\citeauthoryear{{Han}}{{Han}}{2010}]{Han_et_al_2010}
{Han} W.-B.,  2010, \mn@doi [\prd] {10.1103/PhysRevD.82.084013}, \href
  {https://ui.adsabs.harvard.edu/abs/2010PhRvD..82h4013H} {82, 084013}

\bibitem[\protect\citeauthoryear{Hernquist}{Hernquist}{1990}]{Hernquist_1990}
Hernquist L.,  1990, \mn@doi [ApJ] {10.1086/168845}, \href
  {http://adsabs.harvard.edu/abs/1990ApJ...356..359H} {356, 359}

\bibitem[\protect\citeauthoryear{Hills}{Hills}{1975}]{Hills_1975}
Hills J.~G.,  1975, \mn@doi [Nature] {10.1038/254295a0}, \href
  {https://ui.adsabs.harvard.edu/abs/1975Natur.254..295H/abstract} {254, 295}

\bibitem[\protect\citeauthoryear{{Hils} \& {Bender}}{{Hils} \&
  {Bender}}{1995}]{Hils1995}
{Hils} D.,  {Bender} P.~L.,  1995, \mn@doi [\apjl] {10.1086/187876}, \href
  {https://ui.adsabs.harvard.edu/abs/1995ApJ...445L...7H} {445, L7}

\bibitem[\protect\citeauthoryear{{Hjorth} \& {Madsen}}{{Hjorth} \&
  {Madsen}}{1991}]{Hjorth1991}
{Hjorth} J.,  {Madsen} J.,  1991, \mn@doi [\mnras] {10.1093/mnras/253.4.703},
  \href {https://ui.adsabs.harvard.edu/abs/1991MNRAS.253..703H} {253, 703}

\bibitem[\protect\citeauthoryear{{Holley-Bockelmann} \&
  {Khan}}{{Holley-Bockelmann} \& {Khan}}{2015}]{Holley-Bockelmann2015}
{Holley-Bockelmann} K.,  {Khan} F.~M.,  2015, \mn@doi [\apj]
  {10.1088/0004-637X/810/2/139}, \href
  {https://ui.adsabs.harvard.edu/abs/2015ApJ...810..139H} {810, 139}

\bibitem[\protect\citeauthoryear{{Holley-Bockelmann} \&
  {Sigurdsson}}{{Holley-Bockelmann} \&
  {Sigurdsson}}{2006}]{Holley-Bockelmann2006}
{Holley-Bockelmann} K.,  {Sigurdsson} S.,  2006, arXiv e-prints, \href
  {https://ui.adsabs.harvard.edu/abs/2006astro.ph..1520H} {pp
  astro--ph/0601520}

\bibitem[\protect\citeauthoryear{{Huertas-Company} et~al.,}{{Huertas-Company}
  et~al.}{2020}]{Huertas-Company2020}
{Huertas-Company} M.,  et~al., 2020, \mn@doi [\mnras] {10.1093/mnras/staa2777},
  \href {https://ui.adsabs.harvard.edu/abs/2020MNRAS.499..814H} {499, 814}

\bibitem[\protect\citeauthoryear{Jaffe}{Jaffe}{1983}]{Jaffe_1983}
Jaffe W.,  1983, \mn@doi [MNRAS] {10.1093/mnras/202.4.995}, \href
  {http://adsabs.harvard.edu/abs/1983MNRAS.202..995J} {202, 995}

\bibitem[\protect\citeauthoryear{Kashiyama \& Inayoshi}{Kashiyama \&
  Inayoshi}{2016}]{Kashiyama_2016}
Kashiyama K.,  Inayoshi K.,  2016, \mn@doi [ApJ] {10.3847/0004-637X/826/1/80},
  \href {https://ui.adsabs.harvard.edu/abs/2016ApJ...826...80K/abstract} {826,
  80}

\bibitem[\protect\citeauthoryear{K{\i}z{\i}ltan, Baumgardt  \&
  Loeb}{K{\i}z{\i}ltan et~al.}{2017}]{Kiziltan_2017}
K{\i}z{\i}ltan B.,  Baumgardt H.,   Loeb A.,  2017, \mn@doi [Nature]
  {10.1038/nature21361}, \href
  {http://adsabs.harvard.edu/abs/2017Natur.542..203K} {542, 203}

\bibitem[\protect\citeauthoryear{{Kormendy} \& {Ho}}{{Kormendy} \&
  {Ho}}{2013}]{Kormendy2013}
{Kormendy} J.,  {Ho} L.~C.,  2013, \mn@doi [\araa]
  {10.1146/annurev-astro-082708-101811}, \href
  {http://adsabs.harvard.edu/abs/2013ARA%26A..51..511K} {51, 511}

\bibitem[\protect\citeauthoryear{Kroupa}{Kroupa}{2001}]{Kroupa_2001}
Kroupa P.,  2001, \mn@doi [MNRAS] {10.1046/j.1365-8711.2001.04022.x}, \href
  {http://adsabs.harvard.edu/abs/2001MNRAS.322..231K} {322, 231}

\bibitem[\protect\citeauthoryear{{Lake} \& {Norman}}{{Lake} \&
  {Norman}}{1983}]{Lake1983}
{Lake} G.,  {Norman} C.,  1983, \mn@doi [\apj] {10.1086/161097}, \href
  {https://ui.adsabs.harvard.edu/abs/1983ApJ...270...51L} {270, 51}

\bibitem[\protect\citeauthoryear{Lanzoni et~al.,}{Lanzoni
  et~al.}{2013}]{Lanzoni_2013}
Lanzoni B.,  et~al., 2013, \mn@doi [ApJ] {10.1088/0004-637X/769/2/107}, \href
  {http://adsabs.harvard.edu/abs/2013ApJ...769..107L} {769, 107}

\bibitem[\protect\citeauthoryear{{Lezhnin} \& {Vasiliev}}{{Lezhnin} \&
  {Vasiliev}}{2015}]{Lezhnin2015}
{Lezhnin} K.,  {Vasiliev} E.,  2015, \mn@doi [\apjl]
  {10.1088/2041-8205/808/1/L5}, \href
  {https://ui.adsabs.harvard.edu/abs/2015ApJ...808L...5L} {808, L5}

\bibitem[\protect\citeauthoryear{{Lezhnin} \& {Vasiliev}}{{Lezhnin} \&
  {Vasiliev}}{2016}]{Lezhnin2016}
{Lezhnin} K.,  {Vasiliev} E.,  2016, \mn@doi [\apj]
  {10.3847/0004-637X/831/1/84}, \href
  {https://ui.adsabs.harvard.edu/abs/2016ApJ...831...84L} {831, 84}

\bibitem[\protect\citeauthoryear{{Lin} \& {Tremaine}}{{Lin} \&
  {Tremaine}}{1980}]{Lin1980}
{Lin} D.~N.~C.,  {Tremaine} S.,  1980, \mn@doi [\apj] {10.1086/158513}, \href
  {https://ui.adsabs.harvard.edu/abs/1980ApJ...242..789L} {242, 789}

\bibitem[\protect\citeauthoryear{{Lin} et~al.,}{{Lin} et~al.}{2018}]{Lin2018}
{Lin} D.,  et~al., 2018, \mn@doi [Nature Astronomy]
  {10.1038/s41550-018-0493-1}, \href
  {https://ui.adsabs.harvard.edu/abs/2018NatAs...2..656L} {2, 656}

\bibitem[\protect\citeauthoryear{{Luo} et~al.,}{{Luo} et~al.}{2016}]{Luo2016}
{Luo} J.,  et~al., 2016, \mn@doi [Classical and Quantum Gravity]
  {10.1088/0264-9381/33/3/035010}, \href
  {https://ui.adsabs.harvard.edu/abs/2016CQGra..33c5010L} {33, 035010}

\bibitem[\protect\citeauthoryear{L{\"u}tzgendorf, {Kissler-Patig}, Gebhardt,
  Baumgardt, Noyola, Jalali, {de Zeuuw}  \& Neumayer}{L{\"u}tzgendorf
  et~al.}{2012}]{Lutzgendorf_kis_2012}
L{\"u}tzgendorf N.,  {Kissler-Patig} M.,  Gebhardt K.,  Baumgardt H.,  Noyola
  E.,  Jalali B.,  {de Zeuuw} P.~T.,   Neumayer N.,  2012, \mn@doi [Astron.
  Astrophys] {10.1051/0004-6361/201219375}, \href
  {http://adsabs.harvard.edu/abs/2012A\%26A...542A.129L} {542}

\bibitem[\protect\citeauthoryear{L{\"u}tzgendorf et~al.,}{L{\"u}tzgendorf
  et~al.}{2013}]{Lutzgendorf_ki_2013}
L{\"u}tzgendorf N.,  et~al., 2013, \mn@doi [Astron. Astrophys]
  {10.1051/0004-6361/201220307}, \href
  {http://adsabs.harvard.edu/abs/2013A\%26A...552A..49L} {552}

\bibitem[\protect\citeauthoryear{Maccarone \& Servillat}{Maccarone \&
  Servillat}{2008}]{Maccarone_2008}
Maccarone T.~J.,  Servillat M.,  2008, \mn@doi [MNRAS]
  {10.1111/j.1365-2966.2008.13577.x}, \href
  {http://adsabs.harvard.edu/abs/2008MNRAS.389..379M} {389, 379}

\bibitem[\protect\citeauthoryear{Magano, {Vilas Boas}  \& Martins}{Magano
  et~al.}{2017}]{Magano_2017}
Magano D. M.~N.,  {Vilas Boas} J. M.~A.,   Martins C. J. A.~P.,  2017, \mn@doi
  [Phys. Rev. D] {10.1103/PhysRevD.96.083012}, \href
  {https://ui.adsabs.harvard.edu/search/q=author\%3A\%22Martins\%2C\%20C.\%20J.\%20A.\%20P.\%22&sort=date\%20desc\%2C\%20bibcode\%20desc&p_=0}
  {96}

\bibitem[\protect\citeauthoryear{{Mandelker}, {Dekel}, {Ceverino}, {Tweed},
  {Moody}  \& {Primack}}{{Mandelker} et~al.}{2014}]{Mandelker_et_al_2014}
{Mandelker} N.,  {Dekel} A.,  {Ceverino} D.,  {Tweed} D.,  {Moody} C.~E.,
  {Primack} J.,  2014, \mn@doi [\mnras] {10.1093/mnras/stu1340}, \href
  {https://ui.adsabs.harvard.edu/abs/2014MNRAS.443.3675M} {443, 3675}

\bibitem[\protect\citeauthoryear{{Mapelli}}{{Mapelli}}{2018}]{Mapelli2018}
{Mapelli} M.,  2018, arXiv e-prints, \href
  {https://ui.adsabs.harvard.edu/abs/2018arXiv180909130M} {p. arXiv:1809.09130}

\bibitem[\protect\citeauthoryear{Merloni et~al.,}{Merloni
  et~al.}{2012}]{eROSITA_2012}
Merloni A.,  et~al., 2012, eROSITA Science Book: Mapping the Structure of the
  Energetic Universe (\mn@eprint {arXiv} {arXiv:1209.3114v2})

\bibitem[\protect\citeauthoryear{{Merritt}}{{Merritt}}{2001}]{Merritt2001}
{Merritt} D.,  2001, \mn@doi [\apj] {10.1086/321550}, \href
  {https://ui.adsabs.harvard.edu/abs/2001ApJ...556..245M} {556, 245}

\bibitem[\protect\citeauthoryear{{Merritt}}{{Merritt}}{2005}]{Merritt2005}
{Merritt} D.,  2005, \mn@doi [\apj] {10.1086/429398}, \href
  {https://ui.adsabs.harvard.edu/abs/2005ApJ...628..673M} {628, 673}

\bibitem[\protect\citeauthoryear{{Merritt}}{{Merritt}}{2013}]{Merritt_2013}
{Merritt} D.,  2013, {Dynamics and Evolution of Galactic Nuclei}.
Princeton University Press

\bibitem[\protect\citeauthoryear{{Merritt}}{{Merritt}}{2015}]{Merritt2015}
{Merritt} D.,  2015, \mn@doi [\apj] {10.1088/0004-637X/810/1/2}, \href
  {https://ui.adsabs.harvard.edu/abs/2015ApJ...810....2M} {810, 2}

\bibitem[\protect\citeauthoryear{{Merritt} \& {Poon}}{{Merritt} \&
  {Poon}}{2004}]{MerrittPoon2004}
{Merritt} D.,  {Poon} M.~Y.,  2004, \mn@doi [\apj] {10.1086/382497}, \href
  {https://ui.adsabs.harvard.edu/abs/2004ApJ...606..788M} {606, 788}

\bibitem[\protect\citeauthoryear{{Merritt}, {Alexander}, {Mikkola}  \&
  {Will}}{{Merritt} et~al.}{2011}]{Merritt_et_al_2011}
{Merritt} D.,  {Alexander} T.,  {Mikkola} S.,   {Will} C.~M.,  2011, \mn@doi
  [\prd] {10.1103/PhysRevD.84.044024}, \href
  {https://ui.adsabs.harvard.edu/abs/2011PhRvD..84d4024M} {84, 044024}

\bibitem[\protect\citeauthoryear{Mezcua}{Mezcua}{2017}]{Mezcua_2017}
Mezcua M.,  2017, \mn@doi [Int. J. Mod. Phys. A] {10.1142/S021827181730021X},
  \href {https://ui.adsabs.harvard.edu/abs/2017IJMPD..2630021M/abstract} {26,
  1730021}

\bibitem[\protect\citeauthoryear{{Mezcua}, {Civano}, {Marchesi}, {Suh},
  {Fabbiano}  \& {Volonteri}}{{Mezcua} et~al.}{2018}]{Mezcua_et_al_2018}
{Mezcua} M.,  {Civano} F.,  {Marchesi} S.,  {Suh} H.,  {Fabbiano} G.,
  {Volonteri} M.,  2018, \mn@doi [\mnras] {10.1093/mnras/sty1163}, \href
  {https://ui.adsabs.harvard.edu/abs/2018MNRAS.478.2576M} {478, 2576}

\bibitem[\protect\citeauthoryear{Miller \& Hamilton}{Miller \&
  Hamilton}{2002}]{Miller_2002}
Miller M.~C.,  Hamilton D.~P.,  2002, \mn@doi [MNRAS]
  {10.1046/j.1365-8711.2002.05112.x}, \href
  {http://adsabs.harvard.edu/abs/2002MNRAS.330..232C} {330, 232}

\bibitem[\protect\citeauthoryear{{Morawski}, {Giersz}, {Askar}  \&
  {Belczynski}}{{Morawski} et~al.}{2018}]{Morawski2018}
{Morawski} J.,  {Giersz} M.,  {Askar} A.,   {Belczynski} K.,  2018, \mn@doi
  [\mnras] {10.1093/mnras/sty2401}, \href
  {https://ui.adsabs.harvard.edu/abs/2018MNRAS.481.2168M} {481, 2168}

\bibitem[\protect\citeauthoryear{Noyola, Gebhardt, {Kissler-Patig},
  L{\"u}tzgendorf, Jalali, {de Zeeuw}  \& Baumgardt}{Noyola
  et~al.}{2010}]{Noyola_2010}
Noyola E.,  Gebhardt K.,  {Kissler-Patig} M.,  L{\"u}tzgendorf N.,  Jalali B.,
  {de Zeeuw} P.~T.,   Baumgardt H.,  2010, \mn@doi [ApJ]
  {10.1088/2041-8205/719/1/L60}, \href
  {http://adsabs.harvard.edu/abs/2010ApJ...719L..60N} {719, L60}

\bibitem[\protect\citeauthoryear{Perera et~al.,}{Perera
  et~al.}{2017}]{Perera_2017}
Perera B. B.~P.,  et~al., 2017, \mn@doi [MNRAS] {10.1093/mnras/stx501}, \href
  {http://adsabs.harvard.edu/abs/2017MNRAS.468.2114P} {468, 2114}

\bibitem[\protect\citeauthoryear{{Pfister}, {Bar-Or}, {Volonteri}, {Dubois}  \&
  {Capelo}}{{Pfister} et~al.}{2019}]{Pfister_et_al_2019}
{Pfister} H.,  {Bar-Or} B.,  {Volonteri} M.,  {Dubois} Y.,   {Capelo} P.~R.,
  2019, \mn@doi [\mnras] {10.1093/mnrasl/slz091}, \href
  {https://ui.adsabs.harvard.edu/abs/2019MNRAS.488L..29P} {488, L29}

\bibitem[\protect\citeauthoryear{{Pfister}, {Volonteri}, {Dai}  \&
  {Colpi}}{{Pfister} et~al.}{2020}]{Pfister_et_al_2020}
{Pfister} H.,  {Volonteri} M.,  {Dai} J.~L.,   {Colpi} M.,  2020, \mn@doi
  [\mnras] {10.1093/mnras/staa1962}, \href
  {https://ui.adsabs.harvard.edu/abs/2020MNRAS.tmp.2074P} {}

\bibitem[\protect\citeauthoryear{{Planck Collaboration} et~al.,}{{Planck
  Collaboration} et~al.}{2020}]{Planck_et_al_2018}
{Planck Collaboration} et~al., 2020, \mn@doi [\aap]
  {10.1051/0004-6361/201833910}, \href
  {https://ui.adsabs.harvard.edu/abs/2020A&A...641A...6P} {641, A6}

\bibitem[\protect\citeauthoryear{{Portegies Zwart} \& {McMillan}}{{Portegies
  Zwart} \& {McMillan}}{2002}]{Portegies-Zwart2002}
{Portegies Zwart} S.~F.,  {McMillan} S. L.~W.,  2002, \mn@doi [\apj]
  {10.1086/341798}, \href
  {https://ui.adsabs.harvard.edu/abs/2002ApJ...576..899P} {576, 899}

\bibitem[\protect\citeauthoryear{{Portegies Zwart}, {Baumgardt}, {Hut},
  {Makino}  \& {McMillan}}{{Portegies Zwart}
  et~al.}{2004}]{Portegies-Zwart2004}
{Portegies Zwart} S.~F.,  {Baumgardt} H.,  {Hut} P.,  {Makino} J.,   {McMillan}
  S. L.~W.,  2004, \mn@doi [\nat] {10.1038/nature02448}, \href
  {https://ui.adsabs.harvard.edu/abs/2004Natur.428..724P} {428, 724}

\bibitem[\protect\citeauthoryear{{Preto} \& {Amaro-Seoane}}{{Preto} \&
  {Amaro-Seoane}}{2010}]{Preto2010}
{Preto} M.,  {Amaro-Seoane} P.,  2010, \mn@doi [\apjl]
  {10.1088/2041-8205/708/1/L42}, \href
  {https://ui.adsabs.harvard.edu/abs/2010ApJ...708L..42P} {708, L42}

\bibitem[\protect\citeauthoryear{{Rauch} \& {Tremaine}}{{Rauch} \&
  {Tremaine}}{1996}]{Rauch1996}
{Rauch} K.~P.,  {Tremaine} S.,  1996, \mn@doi [\na]
  {10.1016/S1384-1076(96)00012-7}, \href
  {https://ui.adsabs.harvard.edu/abs/1996NewA....1..149R} {1, 149}

\bibitem[\protect\citeauthoryear{Rees}{Rees}{1988}]{Rees_1988}
Rees M.~J.,  1988, \mn@doi [Nature] {10.1038/333523a0}, \href
  {http://adsabs.harvard.edu/abs/1988Natur.333..523R} {333, 523}

\bibitem[\protect\citeauthoryear{{Reines} \& {Volonteri}}{{Reines} \&
  {Volonteri}}{2015}]{Reines_Volonteri_2015}
{Reines} A.~E.,  {Volonteri} M.,  2015, \mn@doi [\apj]
  {10.1088/0004-637X/813/2/82}, \href
  {https://ui.adsabs.harvard.edu/abs/2015ApJ...813...82R} {813, 82}

\bibitem[\protect\citeauthoryear{{Reines}, {Greene}  \& {Geha}}{{Reines}
  et~al.}{2013}]{Reines_et_al_2013}
{Reines} A.~E.,  {Greene} J.~E.,   {Geha} M.,  2013, \mn@doi [\apj]
  {10.1088/0004-637X/775/2/116}, \href
  {https://ui.adsabs.harvard.edu/abs/2013ApJ...775..116R} {775, 116}

\bibitem[\protect\citeauthoryear{{Reinoso}, {Schleicher}, {Fellhauer}, {Leigh}
  \& {Klessen}}{{Reinoso} et~al.}{2020}]{Reinoso2020}
{Reinoso} B.,  {Schleicher} D.~R.~G.,  {Fellhauer} M.,  {Leigh} N.~W.~C.,
  {Klessen} R.~S.,  2020, \mn@doi [\aap] {10.1051/0004-6361/202037843}, \href
  {https://ui.adsabs.harvard.edu/abs/2020A&A...639A..92R} {639, A92}

\bibitem[\protect\citeauthoryear{{Sahu}, {Graham}  \& {Davis}}{{Sahu}
  et~al.}{2019}]{Sahu_et_al_2019}
{Sahu} N.,  {Graham} A.~W.,   {Davis} B.~L.,  2019, \mn@doi [\apj]
  {10.3847/1538-4357/ab0f32}, \href
  {https://ui.adsabs.harvard.edu/abs/2019ApJ...876..155S} {876, 155}

\bibitem[\protect\citeauthoryear{{Shapiro} \& {Teukolsky}}{{Shapiro} \&
  {Teukolsky}}{1983}]{Shapiro_Teukolsky_1983}
{Shapiro} S.~L.,  {Teukolsky} S.~A.,  1983, {Black Holes, White Dwarfs, and
  Neutron Stars: the Physics of Compact Objects}.
Wiley-Interscience

\bibitem[\protect\citeauthoryear{{Shibuya}, {Ouchi}, {Kubo}  \&
  {Harikane}}{{Shibuya} et~al.}{2016}]{Shibuya_et_al_2016}
{Shibuya} T.,  {Ouchi} M.,  {Kubo} M.,   {Harikane} Y.,  2016, \mn@doi [\apj]
  {10.3847/0004-637X/821/2/72}, \href
  {https://ui.adsabs.harvard.edu/abs/2016ApJ...821...72S} {821, 72}

\bibitem[\protect\citeauthoryear{Stadel}{Stadel}{2001}]{Stadel_2001}
Stadel J.,  2001, PhD thesis, University of Washington

\bibitem[\protect\citeauthoryear{{Stegmann}, {Capelo}, {Bortolas}  \&
  {Mayer}}{{Stegmann} et~al.}{2020}]{Stegmann_et_al_2020}
{Stegmann} J.,  {Capelo} P.~R.,  {Bortolas} E.,   {Mayer} L.,  2020, \mn@doi
  [\mnras] {10.1093/mnras/staa170}, \href
  {https://ui.adsabs.harvard.edu/abs/2020MNRAS.492.5247S} {492, 5247}

\bibitem[\protect\citeauthoryear{{Stinson}, {Seth}, {Katz}, {Wadsley},
  {Governato}  \& {Quinn}}{{Stinson} et~al.}{2006}]{Stinson_et_al_2006}
{Stinson} G.,  {Seth} A.,  {Katz} N.,  {Wadsley} J.,  {Governato} F.,   {Quinn}
  T.,  2006, \mn@doi [\mnras] {10.1111/j.1365-2966.2006.11097.x}, \href
  {https://ui.adsabs.harvard.edu/abs/2006MNRAS.373.1074S} {373, 1074}

\bibitem[\protect\citeauthoryear{{Stone}, {Vasiliev}, {Kesden}, {Rossi},
  {Perets}  \& {Amaro-Seoane}}{{Stone} et~al.}{2020}]{Stone2020}
{Stone} N.~C.,  {Vasiliev} E.,  {Kesden} M.,  {Rossi} E.~M.,  {Perets} H.~B.,
  {Amaro-Seoane} P.,  2020, \mn@doi [\ssr] {10.1007/s11214-020-00651-4}, \href
  {https://ui.adsabs.harvard.edu/abs/2020SSRv..216...35S} {216, 35}

\bibitem[\protect\citeauthoryear{Tamburello, Mayer, Shen  \&
  Wadsley}{Tamburello et~al.}{2015}]{Tamburello_et_al_2015}
Tamburello V.,  Mayer L.,  Shen S.,   Wadsley J.,  2015, \mn@doi [MNRAS]
  {10.1093/mnras/stv1695}, \href
  {http://adsabs.harvard.edu/abs/2015MNRAS.453.2490T} {453, 2490}

\bibitem[\protect\citeauthoryear{{Tamburello}, {Capelo}, {Mayer}, {Bellovary}
  \& {Wadsley}}{{Tamburello} et~al.}{2017}]{Tamburello_et_al_2017}
{Tamburello} V.,  {Capelo} P.~R.,  {Mayer} L.,  {Bellovary} J.~M.,   {Wadsley}
  J.~W.,  2017, \mn@doi [\mnras] {10.1093/mnras/stw2561}, \href
  {https://ui.adsabs.harvard.edu/abs/2017MNRAS.464.2952T} {464, 2952}

\bibitem[\protect\citeauthoryear{{Toscani}, {Rossi}  \& {Lodato}}{{Toscani}
  et~al.}{2020}]{Toscani_et_al_2020}
{Toscani} M.,  {Rossi} E.~M.,   {Lodato} G.,  2020, \mn@doi [\mnras]
  {10.1093/mnras/staa2290}, \href
  {https://ui.adsabs.harvard.edu/abs/2020MNRAS.498..507T} {498, 507}

\bibitem[\protect\citeauthoryear{Tremaine, Richstone, Byun, Dressler, Faber,
  Grillmair, Kormendy  \& Lauer}{Tremaine et~al.}{1994}]{Tremaine_1994}
Tremaine S.,  Richstone D.~O.,  Byun Y.~I.,  Dressler A.,  Faber S.~M.,
  Grillmair C.,  Kormendy J.,   Lauer T.~R.,  1994, \mn@doi [Astron. J]
  {10.1086/116883}, \href {http://adsabs.harvard.edu/abs/1994AJ....107..634T}
  {107, 634}

\bibitem[\protect\citeauthoryear{{Treu} \& {Koopmans}}{{Treu} \&
  {Koopmans}}{2002}]{Treu2002}
{Treu} T.,  {Koopmans} L. V.~E.,  2002, \mn@doi [\apj] {10.1086/341216}, \href
  {https://ui.adsabs.harvard.edu/abs/2002ApJ...575...87T} {575, 87}

\bibitem[\protect\citeauthoryear{Ulvestad, Greene  \& Ho}{Ulvestad
  et~al.}{2007}]{Ulvestad_2007}
Ulvestad J.~S.,  Greene J.~E.,   Ho L.~C.,  2007, \mn@doi [ApJ]
  {10.1086/518784}, \href {http://adsabs.harvard.edu/abs/2007ApJ...661L.151U}
  {661, L151}

\bibitem[\protect\citeauthoryear{{Valiante}, {Agarwal}, {Habouzit}  \&
  {Pezzulli}}{{Valiante} et~al.}{2017}]{Valiante_et_al_2017}
{Valiante} R.,  {Agarwal} B.,  {Habouzit} M.,   {Pezzulli} E.,  2017, \mn@doi
  [\pasa] {10.1017/pasa.2017.25}, \href
  {https://ui.adsabs.harvard.edu/abs/2017PASA...34...31V} {34, e031}

\bibitem[\protect\citeauthoryear{Vasiliev}{Vasiliev}{2017}]{Vasiliev_2017}
Vasiliev E.,  2017, \mn@doi [ApJ] {10.3847/1538-4357/aa8cc8}, \href
  {http://adsabs.harvard.edu/abs/2017ApJ...848...10V} {848, 10}

\bibitem[\protect\citeauthoryear{Vasiliev}{Vasiliev}{2019}]{Vasiliev_2019}
Vasiliev E.,  2019, \mn@doi [MNRAS] {10.1093/mnras/sty2672}, \href
  {http://adsabs.harvard.edu/abs/2019MNRAS.482.1525V} {482, 1525}

\bibitem[\protect\citeauthoryear{{Volonteri}}{{Volonteri}}{2010}]{Volonteri2010}
{Volonteri} M.,  2010, \mn@doi [\aapr] {10.1007/s00159-010-0029-x}, \href
  {https://ui.adsabs.harvard.edu/abs/2010A&ARv..18..279V} {18, 279}

\bibitem[\protect\citeauthoryear{Wadsley, Stadel  \& Quinn}{Wadsley
  et~al.}{2004}]{Wadsley_2004}
Wadsley J.~W.,  Stadel J.,   Quinn T.,  2004, \mn@doi [New Astron.]
  {10.1016/j.newast.2003.08.004}, \href
  {http://adsabs.harvard.edu/abs/2004NewA....9..137W} {9, 137}

\bibitem[\protect\citeauthoryear{{Wadsley}, {Keller}  \& {Quinn}}{{Wadsley}
  et~al.}{2017}]{Wadsley_et_al_2017}
{Wadsley} J.~W.,  {Keller} B.~W.,   {Quinn} T.~R.,  2017, \mn@doi [\mnras]
  {10.1093/mnras/stx1643}, \href
  {https://ui.adsabs.harvard.edu/abs/2017MNRAS.471.2357W} {471, 2357}

\bibitem[\protect\citeauthoryear{Wheeler}{Wheeler}{1968}]{Wheeler_1968}
Wheeler J.~A.,  1968, \mn@doi [Am. Sci.] {1968AmSci..56....1W}, \href
  {https://ui.adsabs.harvard.edu/abs/1968AmSci..56....1W/abstract} {56, 1}

\bibitem[\protect\citeauthoryear{{Will}}{{Will}}{2012}]{Will_2012}
{Will} C.~M.,  2012, \mn@doi [Classical and Quantum Gravity]
  {10.1088/0264-9381/29/21/217001}, \href
  {https://ui.adsabs.harvard.edu/abs/2012CQGra..29u7001W} {29, 217001}

\bibitem[\protect\citeauthoryear{{Wright}}{{Wright}}{2006}]{Wright_2006}
{Wright} E.~L.,  2006, \mn@doi [\pasp] {10.1086/510102}, \href
  {https://ui.adsabs.harvard.edu/abs/2006PASP..118.1711W} {118, 1711}

\bibitem[\protect\citeauthoryear{Wyller}{Wyller}{1970}]{Wyller_1970}
Wyller A.~A.,  1970, \mn@doi [ApJ] {10.1086/150445}, \href
  {http://adsabs.harvard.edu/abs/1970ApJ...160..443W} {160, 443}

\bibitem[\protect\citeauthoryear{{Zhu}, {Vasiliev}, {Li}  \& {Jing}}{{Zhu}
  et~al.}{2018}]{Zhu_et_al_2018}
{Zhu} Q.,  {Vasiliev} E.,  {Li} Y.,   {Jing} Y.,  2018, \mn@doi [\mnras]
  {10.1093/mnras/sty079}, \href
  {https://ui.adsabs.harvard.edu/abs/2018MNRAS.476....2Z} {476, 2}

\makeatother
\end{thebibliography}
\normalsize

\bsp 
\label{lastpage}
\end{document}